\documentclass[11pt,a4paper]{article}
\pdfoutput=1

\usepackage[utf8]{inputenc}
\usepackage[T1]{fontenc}
\usepackage[margin=2.0cm]{geometry}
\usepackage[pdfstartview={FitH},bookmarks=false,linktoc=page,colorlinks=true,linkcolor=blue,citecolor=blue,urlcolor=blue]{hyperref}

\usepackage[numbered]{bookmark}
\usepackage{mathtools}
\usepackage{amssymb}
\usepackage{graphicx}
\usepackage[font=small,labelfont=bf]{caption}
\usepackage{authblk}
\usepackage{cite}

\usepackage[warn]{textcomp}

\usepackage{dsfont}
\usepackage{xcolor}
\usepackage[titles]{tocloft}
\usepackage{float}

\numberwithin{equation}{section}
\allowdisplaybreaks

\DeclareMathOperator{\csch}{csch}

\begin{document}

\title{\vspace{-0.5cm}\textbf{Entanglement entropy and Fisher information metric for closed bosonic strings in homogeneous plane wave background}\vspace{0.5cm}}
\author[a,b]{H. Dimov}
\author[a]{S. Mladenov}
\author[a,c]{R. C. Rashkov}
\author[a]{T. Vetsov}
\affil[a]{\textit{Department of Physics, Sofia University,}\authorcr\textit{5 J. Bourchier Blvd., 1164 Sofia, Bulgaria}\vspace{5pt}}

\affil[b]{\textit{The Bogoliubov Laboratory of Theoretical Physics, JINR,}\authorcr\textit{141980 Dubna,
Moscow region, Russia}\vspace{5pt}}

\affil[c]{\textit{Institute for Theoretical Physics, Vienna University of Technology,}\authorcr\textit{Wiedner Hauptstr. 8--10, 1040 Vienna, Austria}\authorcr\vspace{10pt}\texttt{h\_dimov,smladenov,rash,vetsov@phys.uni-sofia.bg}
\authorcr\vspace{10pt}\texttt{rash@hep.itp.tuwien.ac.at}\vspace{0.0cm}}
\date{}
\maketitle
\vspace{-1cm}
\begin{abstract}
We derive the extended renormalized entanglement entropy (EREE) and the Fisher information metric in the case of closed bosonic strings in homogeneous plane wave background. Our investigations are conducted within the framework of Thermo Field Dynamics (TFD). The formalism is also
illustrated on the example of some particular models in condensed matter
physics and the non-equilibrium case for system with dissipations.
\end{abstract}

\vspace{0.4cm}
\textsc{Keywords:} Thermo field dynamics, entanglement entropy, Fisher metric, string theory

\thispagestyle{empty}

\noindent\rule{\linewidth}{0.75pt}
\vspace{-0.8cm}\tableofcontents
\noindent\rule{\linewidth}{0.75pt}

\newpage

\section{Introduction}
Entanglement entropy (EE) is a measure of how much information is stored in a quantum system. One expects that EE is directly related to the degrees of freedom. In this sense it can be used to gain insight into the quantum dynamics of diverse and complex phenomena. In recent years it also became a powerful bridge between string/gravity and condensed matter physics. For example, in holographic systems, the entanglement entropy is encoded in the geometric features of different string backgrounds \cite{Takayanagi:2006, Takayanagi:2009}. This is closely related to the concept of
emergent spacetime in such models \cite{Raamsdonk:2010, Takayanagi:2017, Verlinde:2016}. The progress so
far suggests that one can also study relevant aspects of string theory on microscopic level by making use of thermodynamic and information-theoretic quantities. 
\\
\indent Our interest is focused specifically on the study of quantum entanglement entropy and the Fisher information for particular string model, namely closed bosonic strings on homogeneous plane wave backgrounds. In general it is difficult to calculate the entanglement entropy, especially in quantum field theory on curved spacetime. However, the recent progress in thermo field dynamics \cite{Takahashi:1975, Takahashi:1996, Fano:1957, Prigogine:1973, Arimitsu:1985xm} offers relatively easy and straightforward way of treating quantum states, which facilitates the derivation of the EE and the Fisher matrix for the relevant models considered in this paper.
\\
\indent Thermo field dynamics requires a ``statistical'' state defined in a double Hilbert space, which is a direct product of the original space and an isomorphic copy of it. If one chooses to work in   the energy basis $\{| n \rangle\}$, where $\hat H\,|n\rangle=E_n\,|n\rangle$, $n=0,1,\dots$, then the bases in the double Hilbert space are labelled as $\{|n\rangle\bigotimes|\tilde n\rangle\}=\{|n\rangle\,|\tilde n\rangle\}=\{|n,\tilde n\rangle\}$. The extended states were defined originally by \cite{Fano:1957, Prigogine:1973, Takahashi:1975}:
\begin{equation}
|\Psi\rangle=\frac{1}{Z}\,e^{-\beta \, H/2}\,|I\rangle,\quad |I\rangle=\sum_n |n,\tilde n\rangle\,,
\end{equation}
where $Z=Z(\beta)$ is the partition function. It was shown in \cite{Suzuki:1985} that the extended state $|I\rangle$ is invariant for any orthogonal complete set $\{|\alpha\rangle\}$, $|I\rangle=\sum_n |n,\tilde n\rangle=\sum_\alpha |\alpha,\tilde\alpha\rangle$. Thus the statistical state $|\Psi\rangle$ is independent of the chosen representation. This result is known as ``the general representation theorem'' in TFD. It allows one to use TFD techniques even in the non-equilibrium case. The notion of double Hilbert space is very useful in treating quantum states directly and facilitates the calculation of entanglement entropy of the quantum systems. Although TFD works for arbitrary non-diagonal Hamiltonians the calculations simplify if one is allowed to work only with diagonal Hamiltonians, which is the case we prefer in this study.
\\
\indent Let the Hamiltonian be a bilinear function in creation and annihilation operators. One can diagonalise
it by an appropriate procedure, commonly known as the Bogoliubov transformation \cite{Bogoliubov:1958, Xiao:2009}. It mixes the creation and annihilation operators, but leaves the form of the commutation relations unchanged. In this case operator eigenvalues, calculated with the diagonalized Hamiltonian on the transformed state functions, remain unchanged. Many such examples exist with important applications in  condensed matter physics and string theory.
\\
\indent This paper is structured as follows. In section \ref{sec EETIQH} we consider rather generic case of a system in equilibrium, where one applies TFD techniques to calculate the extended entanglement entropy and the Fisher information metric. In section \ref{sec ECMP} we show that our result is applicable for certain bosonic and fermionic systems, naturally found in condensed matter systems such as superfluidity, superconductivity and spin chains. In section \ref{sec EECBSHPWB} we calculate the EREE and the Fisher metric for a non-trivial example of closed bosonic string theory in a class of curved plane wave backgrounds. In section \ref{sec NEEEDS} we  consider a non-equilibrium case with dissipation and generalize the formula for EREE found in \cite{Nakagawa:2016}. Finally, in section \ref{sec conclusion} we make a short summary of our results.

\section{Entanglement entropy for time-independent quadratic Hamiltonians}\label{sec EETIQH}

\subsection{Extended entanglement entropy for systems in equilibrium}
Let $\left| {{\phi _n}} \right\rangle$ be a complete basis of eigenfunctions of the operator $F$,
\begin{equation}\label{}
F\left| {{\phi _n}} \right\rangle  = {F_n}\,\left| {{\phi _n}} \right\rangle ,\quad \left\langle {{{\phi _m}}}
 \mathrel{\left | {\vphantom {{{\phi _m}} {{\phi _n}}}}
 \right. \kern-\nulldelimiterspace}
 {{{\phi _n}}} \right\rangle  = {\delta _{mn}}\,.
\end{equation}
In general the Hamiltonian in such a basis is non-diagonal,
\begin{equation}\label{non-d H}
H = \sum\limits_{m\,n} {{H_{mn}}\,a_m^\dag \,{a_n}} ,\quad {H_{mn}} = \left\langle {{\phi _m}} \right|H\left| {{\phi _n}} \right\rangle \,,
\end{equation}
where the creation  and annihilation operators satisfy standard commutation relations,
\begin{equation}\label{}
\left[ {{a_m},\,a_n^\dag } \right] = {\delta _{mn}},\quad \left[ {{a_m},\,a_n^{}} \right] = 0,\quad \left[ {a_m^\dag ,\,a_n^\dag } \right] = 0\,.
\end{equation}
If the Hamiltonian is diagonalizable one can write it in the following form \cite{Bogoliubov:1958, Xiao:2009, NNS:2016}\footnote{For more general discussion on quantum quadratic Hamiltonians see the lecture notes \cite{Derez:2016}.}
\begin{equation}\label{diag H}
 H = \sum\limits_{i = 1}^N {{E_i}\,b_i^\dag \,{b_i} + {E_0}} \,,
\end{equation}
where the energy coefficients $E_i$ and the energy $E_0$ of the ground state depend on the matrix elements $H_{mn}$ of the original Hamiltonian (\ref{non-d H}). The new creation and annihilation operators $b_i^\dag$ and $b_i$ satisfy the same commutation relations as the previous operators $a_n^\dag$ and $a_n$:
\begin{equation}\label{}
\left[ {{b_i},\,b_j^\dag } \right] = {\delta _{ij}},\quad \left[ {{b_i},\,b_j^{}} \right] = 0,\quad \left[ {b_i^\dag ,\,b_j^\dag } \right] = 0\,,\quad \,i,\,j = 1, \ldots ,N\,.
\end{equation}
Following \cite{Nakagawa:2016, Hashizume:2013} we can apply TFD techniques to find the EREE for the new system of quasi-particles, described by the Hamiltonian (\ref{diag H}). Consider the excited states $\left| {{n_1}, \ldots ,{n_N}} \right\rangle $, which satisfy the orthonormal relation
\begin{equation}\label{}
\left\langle {{m_1}, \ldots ,{m_N}\left| {{n_1}, \ldots ,{n_N}} \right.} \right\rangle  = \prod\limits_{i = 1}^N {{\delta _{{m_i},\,{n_i}}}} \,.
\end{equation}
One can write the Hamiltonian in matrix form such as\footnote{Let us clarify the notations to avoid unnecessary confusion. If we define $I=\left\{{n_1}, \ldots ,{n_N}\right\}$, then a non-diagonal Hamiltonian can be written in the form
\[
\hat H=\sum_{n_1,\dots,\,n_N,\hfill\atop m_1,\dots,\,m_N}H_{n_1,\dots,\,n_N,\,m_1,\dots,\,m_N} \,\left| n_1,\dots,\,n_N\right\rangle \left\langle m_1,\dots,\,m_N \right|=\sum_{IJ}H_{IJ} \,\left| I \right\rangle \left\langle J \right|\,.
\]
The last expression allows one to write $\hat H$ as a matrix, where $I$ and $J$ run over all possible states, defined by the quantum numbers $n_i$  (for explicit examples see \cite{Hashizume:2013, Nakagawa:2015b}). }
\begin{equation}\label{HamilMatrixStatic}
\hat H = \sum\limits_{\left\{ {{n_i}} \right\} = 0}^\infty  {\left( {\sum\limits_{i = 1}^N {{E_i}\,{n_i} + {E_0}} } \right)} \,\left| {{n_1}, \ldots ,{n_N}} \right\rangle \left\langle {{n_1}, \ldots ,{n_N}} \right|\,,
\end{equation}
where $n_i=b^\dag_i\,b_i$ are the number operators, $\{n_i\}=\{n_i\}^N_{i=1}=n_1,\dots,n_N$.
Once the Hamiltonian assumes diagonal form it is straightforward to compute the relevant statistical quantities. The first one is the partition function $Z$,
\begin{align}\label{statSumEqCase}
Z = T{r_{\{ i\} }}\left( {{e^{ - \beta \, \hat H}}} \right) = \sum\limits_{\left\{ {{\ell _i}} \right\} = 0}^\infty  {\left\langle {\left\{ {{\ell _i}} \right\}} \right|} {e^{ - \beta \,\hat H}}\left| {\left\{ {{\ell _i}} \right\}} \right\rangle  = \prod\limits_{i = 1}^N {\frac{{{e^{ - \beta \,{E_0}}}}}{{1 - {e^{ - \beta \,{E_i}}}}}}  = \prod\limits_{i = 1}^N {\frac{{{e^{ - {K_0}}}}}{{1 - {e^{ - {K_i}}}}}} \,,
\end{align}
where $\langle \left\{ {{\ell _i}} \right\}| = \langle {\ell _1},\,{\ell _2}, \ldots ,\,{\ell _N}| = \langle {\ell _1}|\langle {\ell _2}| \ldots \langle {\ell _N}|$, and $\beta=1/T$, ($k_B=1$). We also introduce the notations $K_0=\beta\,E_0$ and $K_i=\beta\,E_i$, $i=1,\dots,N$, usually called inverse scaled temperatures. The ordinary density matrix in equilibrium is given by
\begin{equation}\label{equilDensityMatrix}
{{\hat \rho }_{eq}} = \frac{{{e^{ - \beta \,\hat H}}}}{Z} = \frac{1}{Z}\,\sum\limits_{\{ {n_i}\}  = 0}^\infty  {{e^{- {\sum\limits_{i = 1}^N {{K_i}\,{n_i} - {K_0}} }}}} \,\left| {\{ {n_i}\} } \right\rangle \langle \{ {n_i}\} |\,.
\end{equation}
In order to define the entanglement entropy the whole system is divided into two subsystems $A$ and $B$, traditionally called ``Alice'' and ``Bob''. Then, the standard EE $\Sigma_A$ for the first system is found as\footnote{We prefer to work in units $k_B=1$, where $k_B$ is the Boltzmann constant.}
\begin{equation}\label{standard EE}
\Sigma_A=-k_B\,\rm{Tr}_A\rho_A\,log\rho_A,\quad \rho_A=\rm{Tr}_B\hat\rho_{eq} \,.
\end{equation}
In the TFD formulation of the double Hilbert space the statistical state, $\left| \Psi \right\rangle$, is defined as
\begin{equation}\label{}
\left| \Psi  \right\rangle  = \sum\limits_{\{ {n_i}\}  = 0}^\infty  {\sqrt {{{\hat \rho }_{eq}}} \,\left| {\{ {n_i}\} } \right\rangle \,\left| {\{ {{\tilde n}_i}\} } \right\rangle }  = \frac{1}{{\sqrt Z }}\,\sum\limits_{\{ {n_i}\}  = 0}^\infty  {{e^{ - \frac{1 }{2}\,\left( {\sum\limits_{i = 1}^N {{K_i}\,{n_i} + {K_0}} } \right)}}} \,\left| {\{ {n_i}\} } \right\rangle \,\left| {\{ {{\tilde n}_i}\} } \right\rangle \,.
\end{equation}
Thus the extended density operator assumes the form
\begin{equation}\label{}
\hat \rho =\left| \Psi  \right\rangle \left\langle \Psi  \right| = \frac{1}{Z}\,\sum\limits_{\{ {n_i}\}  = 0}^\infty  {\sum\limits_{\{ {m_i}\}  = 0}^\infty  {{e^{ - \frac{1 }{2}\, \left( {\sum\limits_{i = 1}^N {{K_i}\,({n_i} + {m_i}) + 2\,{K_0}} } \right)}}} } \,\left| {\{ {n_i}\} } \right\rangle \langle \{ {m_i}\} |\left| {\{ {{\tilde n}_i}\} } \right\rangle \langle \{ {{\tilde m}_i}\} |\,.
\end{equation}
One can choose a bipartite system, namely
\begin{align}\label{bipartitioning}
\left\{ {{n_i}} \right\}_{i = 1}^N = \left\{ {{n_\mu}} \right\}_{\mu  = 1}^p\bigcup {\left\{ {{n_k}} \right\}_{k = p + 1}^N} \,,\quad p \le N - 1,\quad N \ge 2\,.
\end{align}
The extended density matrix $\hat\rho_A$ for ``Alice'' is obtained as a trace over the parameters of the second system $B$,
\begin{align}\label{}
{\hat \rho _A} = T{r_{\{ B\} }}\hat \rho  = \sum\limits_{\{ {\ell _k}\}  = 0}^\infty  {\sum\limits_{\{ {{\tilde \ell }_k}\}  = 0}^\infty  {\langle \{ {\ell _k}\} |\langle \{ {{\tilde \ell }_k}\} |\hat \rho \left| {\left\{ {{\ell _k}} \right\}} \right\rangle |\{ {{\tilde \ell }_k}\} \rangle } } \,,
\end{align}
which leads to
\begin{equation}\label{}
{\hat \rho _A} = \sum\limits_{\left\{ {{n_\mu }} \right\} = 0}^\infty  {\sum\limits_{\left\{ {{m_\mu }} \right\} = 0}^\infty  {{e^{ - \frac{1}{2}\,\sum\limits_{\mu  = 1}^p {{K_{{\mu}}}\,\left( {2 + {n_\mu } + {m_\mu }} \right)} }}} } {\,} \left| {\{ {n_\mu }\} } \right\rangle \langle \{ {m_\mu }\} |\left| {\{ {{\tilde n}_\mu }\} } \right\rangle \langle \{ {\tilde m_\mu }\} |\,\prod\limits_{\alpha  = 1}^p {\left( {{e^{{K_\alpha }}} - 1} \right)} \,.
\end{equation}
Finally, the extended renormalized entanglement entropy, ${S_A} =  -\,T{r_{\{ A\} }}\left( {{{\hat \rho }_A}\,\ln {{\hat \rho }_A}} \right)$, follows as
\begin{align}\label{EE generalExp}
{S_A}\left( {{K_\mu }} \right) =  - \sum\limits_{\mu  = 1}^p {\left\{ {\ln \left( {{e^{{K_{{\mu}}}}} - 1} \right) - {K_\mu } - \frac{{{K_\mu }\,\prod\limits_{\gamma  \ne \mu }^{} {\left( {{e^{{K_\gamma }/2}} - 1} \right)} }}{{\prod\limits_{\alpha  = 1}^p {\left( {{e^{{K_\alpha }/2}} - 1} \right)} }}} \right\}} \,\prod\limits_{\alpha  = 1}^p {\coth \frac{{{K_\alpha }}}{4}} \,.
\end{align}
The result simplifies in terms of hyperbolic functions:
\begin{equation}\label{EE general}
{S_A}\left( {{K_\mu }} \right) = \frac{{1}}{2}\,\left( {\prod\limits_{\mu  = 1}^p {\coth \frac{{{K_\mu }}}{4}} } \right)\,\sum\limits_{\mu  = 1}^p {\left\{ {{K_\mu }\,\left( {1 + \coth \frac{{{K_\mu }}}{4}} \right) - 2\,\ln \left( {{e^{{K_\mu }}} - 1} \right)} \right\}} \,.
\end{equation}
This is the desired expression for the EREE. If $p=1,$ the formula reduces to (\ref{EE p1}). If $p=2$, it reproduces the result for the EE of the Pais-Uhlenbeck oscillator, found in \cite{DMRV:2016}. For comparison the standard entanglement entropy from (\ref{standard EE}) is written by
\begin{equation}\label{standard EE explicit}
{{\Sigma}_A}(K_\mu) = \sum\limits_{\mu  = 1}^p {\left[ {\frac{{{K_\mu }}}{4}\,\prod\limits_{\gamma  \ne \mu }^{} {\left( {{e^{{K_\gamma }}} - 1} \right)\,\prod\limits_{\alpha  = 1}^p {\left\{ {\left( {1 - {e^{ - {K_\alpha }}}} \right)\,\csch^2\left( {\frac{{{K_\alpha }}}{2}} \right)} \right\}} }  - \ln \left( {1 - {e^{{-K_\mu }}}} \right)} \right]} \,.
\end{equation}
\subsection{Fisher information metric}

Equation (\ref{EE general}) allows one to calculate the Fisher information metric. It  can be expressed as a second derivative of the entanglement entropy \cite{Matsueda:2014a, Matsueda:2014b}:
\begin{align}\label{Fisher matrix}
{g_{\mu \nu }} = {\partial _\mu }{\partial _\nu }S_A =  - \frac{{{1}}}{8}\,F\,\left( {{A_\mu }\,{B_\nu } + {A_\nu }\,{B_\mu } + {C_{\mu \nu }} + E\,{D_{\mu \nu }}} \right)\,,
\end{align}
where $\partial_\mu=\partial/\partial K_\mu$ and
\begin{equation}\label{}
{A_\mu } = 2\,{\rm{csch}}\frac{{{K_\mu }}}{2}\,,
\end{equation}
\begin{equation}\label{}
{B_\mu } = 1 + \coth \frac{{{K_\mu }}}{4} - \frac{{{K_\mu }}}{4}\,{\rm{csc}}{{\rm{h}}^2}\frac{{{K_\mu }}}{4} - \frac{2}{{1 - {e^{ - {K_\mu }}}}}\,,
\end{equation}
\begin{equation}\label{}
{C_{\mu \nu }} = {\delta _{\mu \nu }}\,\left[ {\left( {2 - \frac{{{K_\mu }}}{2}\,\coth \frac{{{K_\mu }}}{4}} \right)\,{\rm{csc}}{{\rm{h}}^2}\frac{{{K_\mu }}}{4} + \frac{4}{{1 - \cosh {K_\mu }}}} \right]\,,
\end{equation}
\begin{equation}\label{}
{D_{\mu \nu }} = 2\,{\rm{csc}}{{\rm{h}}^2}\frac{{{K_\nu }}}{4}\,\left( {{\delta _{\mu \nu }} + \tanh \frac{{{K_\nu }}}{4}\,\sum\limits_{\tau  \ne \nu } {\left\{ {{\delta _{\mu \tau }}\,{\rm{csch}}\frac{{{K_\tau }}}{2}} \right\}} } \right)\,,
\end{equation}
\begin{equation}\label{}
E =  - \frac{1}{4}\,\sum\limits_{\alpha  = 1}^p {\left[ {{K_\alpha }\,\left( {1 + \coth \frac{{{K_\alpha }}}{4}} \right) - 2\,\ln \left( {{e^{{K_\alpha }}} - 1} \right)} \right]} \,,
\end{equation}
\begin{equation}\label{}
F = \prod\limits_{\sigma  = 1}^p {\coth \frac{{{K_\sigma }}}{4}} \,.
\end{equation}
\indent Formula (\ref{Fisher matrix}) differs by a sign from the standard definition of the metric due to the requirement that the metric components be positive defined, which is a necessary condition for thermodynamic stability (for extended discussion see \cite{GR:1995} and references therein). The case of $p=2$ corresponds to the Fisher metric obtained by \cite{DMRV:2016, DMRV:2017}. 
\\
\indent On the level of the space of probability distributions the Fisher metric represents a continuous setting even if
the underlying features of the system (for example, the state space) are discrete. This allows one to take advantage of the powerful framework
of differential geometry to treat statistical structures as geometrical
ones. As it turns out the expressions for the EE and Fisher metric are applicable for variety of system as shown below.

\section{Examples from condensed matter physics}\label{sec ECMP}

Diagonalizable or approximately diagonalizable bosonic and fermionic Hamiltonians naturally arise in condensed matter physics such as spin wave theory, Heisenberg ferro- and anti-ferromagnets, spin chains, spin liquids, BCS theory of superconductivity \cite{BCS:1957}, but also in quantum field theory and string theory. In this section we give explicit examples of EREE for bosonic and fermionic systems, correspondingly.

\subsection{Entanglement entropy for bosonic system}\label{secBosons}

\begin{figure}[H]
\begin{center}
\includegraphics[scale=1.0]{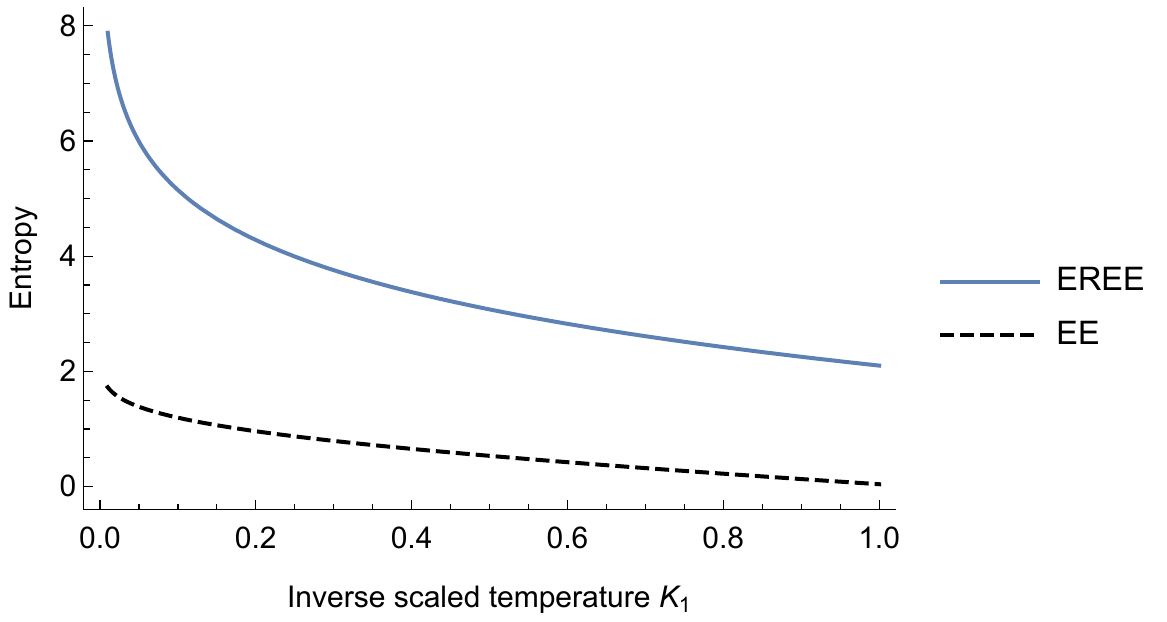}
\caption{The extended renormalized entanglement entropy $S_1$ (thick line) compared to the standard entanglement entropy $\Sigma_1$ (dashed line). As expected the EREE is bigger than the normal EE, but both diverge at the origin (at very high-temperatures).\label{figEREEtoEE}}
\end{center}
\end{figure}

\begin{figure}[H]
\begin{center}
\includegraphics[scale=1]{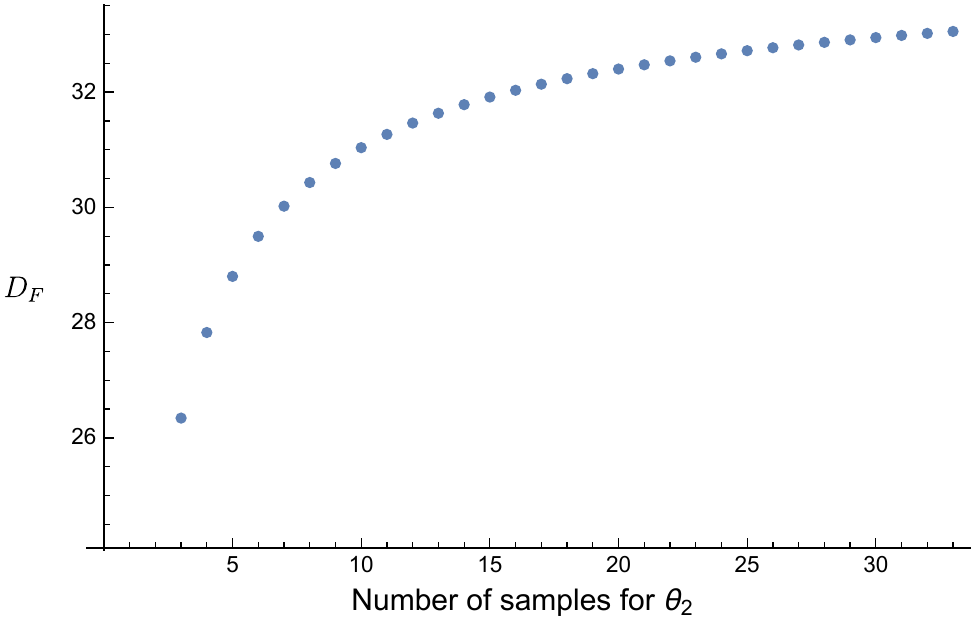}
\caption{Monotonously increasing Fisher information distance $D_F$ between two distributions $f(\theta_1, x)$ and $f(\theta_2,x)$ in a single parameter family for $\theta_1=0.1$ and $\theta_2\in[0.3,10]$, with step size $\delta\theta_2=0.3$. One can interpret this within the framework of fluctuation theory as follow: the less the probability of a fluctuation between two states, the further apart they are.\label{figFisherDistance}}
\end{center}
\end{figure}

For simplicity let us consider the following BCS type bosonic Hamiltonian:
\begin{equation}\label{nondiag H}
H = A\,a_1^\dag \,{a_1} + B\,a_2^\dag \,{a_2} + C\,\left( {a_1^\dag \,a_2^\dag  + {a_1}\,{a_2}} \right)\,,
\end{equation}
where $A$, $B$ and $C$ are some energy coefficients and
\begin{equation}\label{com a}
\left[ {{a_i},\,a_j^\dag } \right] = {\delta _{ij}},\quad \left[ {{a_i},\,{a_j}} \right] = 0,\quad i,j = 1,2\,.
\end{equation}
Following \cite{Bogoliubov:1958}, we want to transform the given Hamiltonian by introducing new set of operators $b^{\dagger}_i$ and $b_i$, such that (\ref{nondiag H}) takes the following diagonal form:
\begin{equation}\label{diag ham}
H = {E_0} + {E_1}\,b_1^\dag \,{b_1} + {E_2}\,b_2^\dag \,{b_2}\,.
\end{equation}
Here, the creation and annihilation operators $b^{\dagger}_i$ and $b_i$ also satisfy (\ref{com a}),
\begin{equation}\label{com b}
\left[ {{b_i},\,b_j^\dag } \right] = {\delta _{ij}},\quad \left[ {{b_i},\,{b_j}} \right] = 0,\quad i,j = 1,2\,.
\end{equation}
The diagonalization is achieved by the following Bogoliubov transformations:
\begin{align}\label{b transf}
\nonumber {b_1} &= \cosh \varphi \,{a_1} + \sinh \varphi \,a_2^\dag \,,\\
{b_2} &= \sinh \varphi \,a_1^\dag  + \cosh \varphi \,{a_2}\,.
\end{align}
After some trivial calculations one arrives at the following expressions for the new Hamiltonian coefficients:
\begin{equation}\label{}
  {E_1} = \frac{1}{2}\,\left( {A - B + \sqrt {{{\left( {A + B} \right)}^2} - 4\,{C^2}} } \right)\,,
\end{equation}
\begin{equation}\label{}
{E_2} = \frac{1}{2}\,\left( {B - A + \sqrt {{{\left( {A + B} \right)}^2} - 4\,{C^2}} } \right)\,,
\end{equation}
\begin{equation}\label{}
{E_0} =  - \left( {{E_1} + {E_2}} \right)\,{\sinh ^2}\varphi \,,
\end{equation}
where $E_0$ is the energy of the ground state.
Now, let us focus on the calculation of the EREE for the system described by the Hamiltonian (\ref{diag ham}). First we calculate the partition function
\begin{equation}\label{}
Z = T{r_{1,{\kern 1pt} 2}}\left( {{e^{ - \beta \,H}}} \right) = \frac{{{e^{ - \beta \,{E_0}}}}}{{\left( {1 - {e^{ - \beta \,{E_1}}}} \right)\,\left( {1 - {e^{ - \beta \,{E_2}}}} \right)}}\,.
\end{equation}
The ordinary equilibrium density matrix is written by
\begin{equation}\label{}
{{\hat \rho }_{eq}} = \frac{{{e^{ - \beta \,H}}}}{Z} = \sum\limits_{{n_1} = 0}^\infty  {\sum\limits_{{n_2} = 0}^\infty  {{e^{ - \beta \,\left( {{E_1}\,{n_1} + {E_2}\,{n_2} + {E_0}} \right)}}} } \left| {{n_1},\,{n_2}} \right\rangle \langle {n_1},\,{n_2}|\,,
\end{equation}
where $n_i=b^{\dagger}_i \,b_i$, $i=1,2$, are the number operators of the Bogoliubov quasi-particles. The TFD statistical state, $\left| \Psi \right\rangle$, is defined as
\begin{equation}\label{}
\left| \Psi  \right\rangle  = \sum\limits_{{n_1} = 0}^\infty  {\sum\limits_{{n_2} = 0}^\infty  {\sqrt {{{\hat \rho }_{eq}}} \,\left| {{n_1},\,{n_2}} \right\rangle \,\left| {{{\tilde n}_1},\,{{\tilde n}_2}} \right\rangle } }  = \frac{1}{{\sqrt Z }}\,\sum\limits_{{n_1} = 0}^\infty  {\sum\limits_{{n_2} = 0}^\infty  {{e^{ - \frac{\beta }{2}\,\left( {{E_1}\,{n_1} + {E_2}\,{n_2} + {E_0}} \right)}}} \,} \left| {{n_1},\,{n_2}} \right\rangle \,\left| {{{\tilde n}_1},\,{{\tilde n}_2}} \right\rangle \,.
\end{equation}
Therefore, the extended density operator, $\hat \rho=\left| \Psi  \right\rangle \left\langle \Psi  \right|$, takes the form
\begin{equation}\label{}
\hat \rho  = \frac{1}{Z}\,\sum\limits_{{n_1} = 0}^\infty  {\sum\limits_{{n_2} = 0}^\infty  {\sum\limits_{{m_1} = 0}^\infty  {\sum\limits_{{m_2} = 0}^\infty  {{e^{ - \frac{\beta }{2}{\kern 1pt} \,\left( {2\,{E_0} + {E_1}\,({n_1} + {m_1}) + {E_2}\,({n_2} + {m_2})} \right)}}} } } } \left| {{n_1},\,{n_2}} \right\rangle \langle {m_1},\,{m_2}|\left| {{{\tilde n}_1},\,{{\tilde n}_2}} \right\rangle \langle {{\tilde m}_1},\,{{\tilde m}_2}|{\mkern 1mu} \,.
\end{equation}
Tracing out the states of the second system, one finds
\begin{equation}\label{}
{{\hat \rho }_1} = T{r_2}(\hat \rho ) = \frac{1}{Z}\,\sum\limits_{{n_1} = 0}^\infty  {\sum\limits_{{m_1} = 0}^\infty  {\frac{{{{\rm{e}}^{ - \beta \,\left( {{E_0} - {E_2} + \frac{{{E_1}\,\left( {{m_1} + {n_1}} \right)}}{2}} \right)\,}}}}{{{{\rm{e}}^{\beta {\kern 1pt} {E_2}}} - 1}}} \,} \left| {{n_1}} \right\rangle \langle {m_1}|\left| {{{\tilde n}_1}} \right\rangle \langle {{\tilde m}_1}|{\mkern 1mu} \,.
\end{equation}
Finally, the renormalized extended entanglement entropy for the given bosonic system is written by
\begin{align}\label{EE p1}
{S_1}({K_1}) =  - T{r_1}\left( {{{\hat \rho }_1}\,\ln {{\hat \rho }_1}} \right) = \frac{{{1}}}{2}\,\coth \frac{{{K_1}}}{4}\,\left\{ {\left( {1 + \coth \frac{{{K_1}}}{4}} \right)\,{K_1} - 2\,\log \left( {{e^{{K_1}}} - 1} \right)} \right\}\,.
\end{align}
Here $K_1=\beta\,E_1$ is the inverse scaled temperature. As expected the result agrees with eq. (\ref{EE general}) for $p=1$. The dependence of the entropy on $K_1$ is illustrated on figure \ref{figEREEtoEE}.
\indent In this case the Fisher information (\ref{Fisher matrix}) is a one parameter function given by
\begin{align}\label{FI function}
\nonumber F({K_1}) &= \frac{{{K_1}}}{{32}}\,{{\mathop{\rm csch}\nolimits} ^4}\frac{{{K_1}}}{4}\,\left( {4 + 2\,\cosh \frac{{{K_1}}}{2} + \sinh \frac{{{K_1}}}{2}} \right)\\
 &- \frac{1}{{16}}\,{{\mathop{\rm csch}\nolimits} ^3}\frac{{{K_1}}}{4}\,\left[ {3 + \log \left( {{e^{{K_1}}} - 1} \right) + \cosh \frac{{{K_1}}}{2}\,\left( {2 + \log \left( {{e^{{K_1}}} - 1} \right)} \right)} \right]\,{\mathop{\rm sech}\nolimits} \frac{{{K_1}}}{4}\,.
\end{align}
It measures the amount of
information that an observed random variable provides about
an unknown parameter. It can be used in studying phase transitions, especially
the second-order phase transitions, during which the Fisher
information exhibits divergence. From eq. (\ref{FI function}) one notices that  Fisher information is singular at the origin $K_1=0$. This suggest that at  very high temperatures the Bogoliubov quasi-system undergoes a second-order phase transition, which is in agreement with the statement that the Fisher information is maximized at the phase transition points \cite{SQHS:2015}.
\\
\indent One can use the Fisher information (\ref{FI function}) to define a distance between points on the statistical manifold, spanned by the inverse scaled temperatures $K_\mu$, or in this case -- only by $\theta=K_1$. The information-metric distance, or Fisher information distance \cite{CSS:2012}, $D_F$, between two distributions $f(\theta_1, x)$ and $f(\theta_2,x)$ in a single parameter family is defined by
\begin{equation}
{D_F}({\theta _1},\,{\theta _2}) = \int\limits_{{\theta _1}}^{{\theta _2}} {\sqrt {F(\theta )} \,d\theta } \,,
\end{equation}
where $\theta_1$ and $\theta_2$ are parameter values corresponding to the
two PDFs. On figure \ref{figFisherDistance} are depicted several values of the Fisher distance $D_F$ for several increasing positive values of the upper integral limit $\theta_2$, while keeping the lower limit $\theta_1$ fixed. This setup chooses different points on the statistical manifolds. One notices that $D_F$ increases monotonously for increasing values of the upper limit $\theta_2$. For nearby states, the square of the lengths of the geodesic paths gives the probability of a fluctuation between the states. In other words, the less the probability of a fluctuation between two states, the further apart they are \cite{GR:1995}.
\subsection{Entanglement entropy for fermionic system}
In this section we are going to consider    a fermionic example, namely XY model in a magnetic field. It is a generalization of the Ising model in which an anisotropy is introduced with respect to the x and y directions by means of a real deformation parameter $\gamma$. In what follows we are going to shortly sketch the diagonalization of the Hamiltonian, which is given by \cite{JDV:2017}:
\begin{equation}
H =  - \sum\limits_{\ell  =  - M}^M {\left[ {\left( {\frac{{1 + \gamma }}{2}} \right)\,\sigma _\ell ^x\,\sigma _{\ell  + 1}^x + \left( {\frac{{1 - \gamma }}{2}} \right)\,\sigma _\ell ^y\,\sigma _{\ell  + 1}^y + h\,\sigma _\ell ^z} \right]} \,.
\end{equation}
Here $N=2\,M+1$ gives the total odd number of spins and $h$ is the transverse magnetic field. In the
 $\gamma= 1$ case the system reduces to the one-dimensional Ising model with transverse magnetic field. In order to diagonalize the Hamiltonian we begin by defining the following operators:
\begin{equation}
{\sigma ^ + } = \frac{1}{2}\,({\sigma ^x} + i\,{\sigma ^y})\,,\quad {\sigma ^ - } = \frac{1}{2}\,({\sigma ^x} - i\,{\sigma ^y})\,.
\end{equation}
Next we perform the Jordan-Wigner transformation, which relates the spin
operators $\sigma_\ell$ to a set of fermionic operators $a_\ell$ and $a^{\dagger}_\ell$ via
\begin{equation}
\sigma _\ell ^ +  = \left( {\prod\limits_{j = 1}^{\ell  - 1} {\sigma _j^z} } \right)\,{a_\ell }\,,\quad \sigma _\ell ^ -  = \left( {\prod\limits_{j = 1}^{\ell  - 1} {\sigma _j^z} } \right)\,a_\ell ^\dag \,,\quad \sigma _\ell ^z = 1 - 2\,a_\ell ^\dag \,{a_\ell }\,.
\end{equation}
Here the operators $a_\ell$ and $a^{\dagger}_\ell$ satisfy the standard fermionic anticommutation relations:
\begin{equation}
\left\{ {a_\ell ^\dag ,\,{a_m}} \right\} = {\delta _{\ell m}}\,,\quad \left\{ {a_\ell ^\dag ,\,a_m^\dag } \right\} = \left\{ {a_\ell ^{},\,a_m^{}} \right\} = 0\,.
\end{equation}
The Hamiltonian, written in terms of these fermionic operators, assumes the form
\begin{align}
\nonumber
H &=  - \sum\limits_{\ell  =  - M}^M {\left[ {\frac{{1 + \gamma }}{2}\,\left( {{a_{\ell  + 1}}\,{a_\ell } + a_{\ell  + 1}^\dag \,{a_\ell } + a_\ell ^\dag \,{a_{\ell  + 1}} + a_\ell ^\dag \,a_{\ell  + 1}^\dag } \right)} \right.} \\
&\left. { + \frac{{\gamma  - 1}}{2}\,\left( {{a_{\ell  + 1}}\,{a_\ell } - a_\ell ^\dag \,{a_{\ell  + 1}} - a_\ell ^\dag \,{a_{\ell  + 1}} + a_\ell ^\dag \,a_{\ell  + 1}^\dag } \right) + h\,\left( {1 - 2\,a_\ell ^\dag \,{a_\ell }} \right)} \right]\,.
\end{align}
Now we Fourier transform the creation and annihilation operators by
\begin{equation}
{a_\ell } = \frac{1}{{\sqrt N }}\,\sum\limits_k {{e^{ - i\,k\,\ell }}\,{d_k}} \,,\quad a_\ell ^\dag  = \frac{1}{{\sqrt N }}\,\sum\limits_k {{e^{i\,k\,\ell }}\,d_k^\dag } \,,\quad {\delta _{kk'}} = \frac{1}{N}\,\sum\limits_\ell  {{e^{i\,\ell \,(k - k')}}} \,,
\end{equation}
where $k=2\,\pi/N,\,4\,\pi/N\,\dots,2\,\pi$. The Hamiltonian is expressed as
\begin{equation}
H =  - \sum\limits_k {\left[ {2\,(\cos k - h)\,d_k^\dag \,{d_k} - i\,\gamma \,\sin k\,({d_k}{d_{ - k}} + d_k^\dag d_{ - k}^\dag )} \right]}  - h\,N\,.
\end{equation}
After applying the following Bogoliubov transformations:
\begin{equation}
{d_k} = \cos \frac{{{\theta _k}}}{2}\,{b_k} + i\,\sin \frac{{{\theta _k}}}{2}\,b_{ - k}^\dag \,,\quad d_k^\dag  = \cos \frac{{{\theta _k}}}{2}\,b_k^\dag  - i\,\sin \frac{{{\theta _k}}}{2}\,{b_{ - k}}\,,
\end{equation}
one finds
\begin{align}
\nonumber
H &=  - \sum\limits_k {2\,\gamma \,\sin k\,\sin {\theta _k}\,b_k^\dag \,{b_k} - 2\,\cos {\theta _k}(\cos k - h)\,b_k^\dag \,{b_k}} \\
 &- i\sum\limits_k {\left( {\sin {\theta _k}(\cos k - h) - \gamma \,\sin k\,\sin {\theta _k}} \right)\,({b_k}{b_{ - k}} + b_k^\dag b_{ - k}^\dag )}  + const\,.
\end{align}
Imposing that the cross-terms be zero we arrive at the expressions relating $\theta_k$ with the original parameters ($k,h,\gamma$):
\begin{equation}
\cos {\theta _k} = \frac{{\cos k - h}}{{\sqrt {{{(\cos k - h)}^2} + {\gamma ^2}\,{{\sin }^2}k} }}\,,\quad \sin {\theta _k} =  - \frac{{\gamma \,\sin k}}{{\sqrt {{{(\cos k - h)}^2} + {\gamma ^2}\,{{\sin }^2}k} }}\,.
\end{equation}
Finally, the Hamiltonian assumes the form of a quasi-free fermionic system:
\begin{equation}
H = \sum\limits_k {{\Lambda _k}\,\left( {n_k - 1} \right)} \,,
\end{equation}
where ${n_k} = b_k^\dag \,{b_k}$ defines number operators for the quasi-particles and $\Lambda_k$ sets the following  dispersion relation:
\begin{equation}
{\Lambda _k} = \sqrt {{{(\cos k - h)}^2} + {\gamma ^2}\,{{\sin }^2}k} \,.
\end{equation}
\indent One can repeat the TFD analysis from section \ref{sec EETIQH} to calculate the extended entanglement entropy, which, for arbitrary number of spins $N$, coincides with equation (\ref{EE general}):
\begin{equation}\label{}
{S_A}\left( {{K_\mu }} \right) = \frac{{1}}{2}\,\left( {\prod\limits_{\mu  = 1}^p {\coth \frac{{{K_\mu }}}{4}} } \right)\,\sum\limits_{\mu  = 1}^p {\left\{ {{K_\mu }\,\left( {1 + \coth \frac{{{K_\mu }}}{4}} \right) - 2\,\ln \left( {{e^{{K_\mu }}} - 1} \right)} \right\}} \,,
\end{equation}
where the inverse scaled temperatures are given by $K_\mu=\beta\,\Lambda_\mu$. Here we used natural numbers to count the number of spins involved. In such notations one has to be careful with the expressions for $\Lambda_\mu$ and $\Lambda_k$, where $1\leq\mu\leq p\leq N-1,\,N>1$, and the angle $k=2\,\pi/p,\,4\,\pi/p\,\dots,2\,\pi$. This implies the following relation $\mu\to k=2\,\pi/\mu$, thus
\begin{equation}
{\Lambda _\mu } \to {\Lambda _{2\,\pi /\mu }} = \sqrt {{{\left( {\cos \left( {\frac{{2\,\pi }}{\mu }} \right) - h} \right)}^2} + {\gamma ^2}\,{{\sin }^2}\left( {\frac{{2\,\pi }}{\mu }} \right)} \,.
\end{equation}
\indent The Fisher information metric in this case is the same as in section \ref{sec EETIQH} for particular values of $k$.
\section{Entanglement entropy for closed bosonic strings in homogeneous plane wave backgrounds}\label{sec EECBSHPWB}
In this section we consider the closed bosonic string vibrating in regular homogeneous plane-wave backgrounds. The given curved backgrounds have non-vanishing NS three-form field strength and a dilaton. We will closely follow \cite{BLPT:2003}, where the authors develop a general procedure for solving linear, but non-diagonal equations for the string coordinates,
and determine the corresponding oscillator frequencies and the light-cone Hamiltonian. In this set up the Hamiltonian is automatically diagonalized and time-independent. Therefore, finding the entanglement entropy in the framework of TFD naturally follows the steps shown in the previous sections. Bellow we will  sketch the relevant result of \cite{BLPT:2003}.
\subsection{String equations of motion and quantization}
We begin by considering a closed relativistic string in non-singular $2 + d$ dimensional homogeneous plane-wave backgrounds with metric of the following form
\begin{equation}
d{s^2} = 2\,du\,dv + {k_{ij}}\,{x^i}\,{x^j}\,d{u^2} + 2\,{f_{ij}}\,{x^i}\,d{x^j}\,du + d{x^i}\,d{x^j}\,.
\end{equation}
Here $k_{ij}$ and $f_{ij}$ are constant, and the $B$-field is given by ${B_{iu}} =  - {h_{ij}}\,{x^j}$. Our aim is to solve the classical equations of motion for this string sigma model. We  denote the string embedding coordinates as $X^M = (U, V,X^i)$. Choosing the orthogonal gauge for the world-sheet
metric, the standard sigma model Lagrangian is written by
\begin{equation}
L = \frac{1}{{2\,\pi }}\,\left( {{G_{MN}}(X) + {B_{MN}}(X)} \right)\,{\partial _ + }{X^M}\,{\partial _ - }{X^N}\,.
\end{equation}
The equations of motion for the bosonic field $U$ are easily obtained:
\begin{equation}
{\partial _ + }{\partial _ - }U = 0\,.
\end{equation}
Similarly, for the fields $X^i$, $i=1,\dots,d$, one finds
\begin{equation}\label{transverse coords}
 - {\partial _ + }{\partial _ - }{X_i} + ({f_{ij}} + {h_{ij}})\,{\partial _ - }U\,{\partial _ + }{X^j} + ({f_{ij}} - {h_{ij}})\,{\partial _ + }U\,{\partial _ - }{X^j} + {k_{ij}}\,{X^j}\,{\partial _ + }U\,{\partial _ - }U = 0\,,
\end{equation}
where ${\sigma ^ \pm } = \tau  \pm \sigma$ and  ${\partial _ \pm } = {\partial _\tau } \pm {\partial _\sigma }$. In the light-cone gauge
$U$ becomes
\begin{equation}
U = {p_ + }\,{\sigma ^ + } + {p_ - }\,{\sigma ^ - } = \frac{{{p_v}}}{2}\,,
\end{equation}
where the condition of periodicity of $U$ in $\sigma$ implies that
${p_ + } = {p_ - } = {p_v}/2$. To solve eq. (\ref{transverse coords}) one makes the following mode expansion of the transverse coordinates
\begin{equation}
{X^i}(\tau ,\,\sigma ) = \sum\limits_{n =  - \infty }^\infty  {X_n^i(\tau )\,{e^{2\,i\,n\,\sigma }}} ,\,\quad X_n^i = {\left( {X_{ - n}^i} \right)^*}\,,\quad 0 < \sigma  \le \pi\,.
\end{equation}
The substitution of the mode expansion in eq. (\ref{transverse coords}) leads to
\begin{equation}\label{eom transv}
 - \ddot X_n^i + 2\,{p_v}\,{f_{ij}}\,\dot X_n^j + (p_v^2\,{k_{ij}} - 4\,{n^2}\,{\delta _{ij}})\,X_n^j + 4\,i\,n\,{p_v}\,{h_{ij}}\,X_n^j = 0\,.
\end{equation}
For simplicity one can set $p_v=1$ and assume that $k_{ij}$ is diagonal,
$k_{ij} = k_i\,\delta_{ij}$. As explained in \cite{BLPT:2003}, the general method to solve  systems like (\ref{eom transv}) is to rewrite it as a set of $2\,d$ first-order equations and then use the appropriate methods available at hand. Fortunately, the authors noted that for generic values of the parameters in (\ref{eom transv}) one can use a much
simpler procedure. Namely, to solve these equations, one makes the following ansatz:
\begin{equation}\label{freq base ans}
X_n^i(\tau) = \sum\limits_{J = 1}^{2\,d} {\zeta _J^{(n)}\,a_{iJ}^{(n)}\,{e^{i\,\omega _J^{(n)}\,\tau }}\,} ,
\end{equation}
with the frequencies $\omega^{(n)}_
J$ and their eigen-directions $a^{(n)}
_{iJ}$ to be determined. This frequency based ansatz for the modes leads to the matrix equation in the form
\begin{equation}\label{the matrix eq}
M_{ik}\left(\omega^{(n)}_J,n\right)\,a^{(n)}_{kJ}=0 \,,
\end{equation}
where (for short $\omega=\omega^{(n)}_J$):
\begin{equation}\label{The M matrix}
M_{ik}=\left(\omega^2+k_i-4\,n^2\right)\,\delta_{ik}+2\,i\,\omega\,f_{ik}+4 \,i\,n\,h_{ik}.
\end{equation}
The matrix equation (\ref{the matrix eq}) is a homogeneous algebraic system. The necessary condition for finding a non-trivial solution is
\begin{equation}
\det{M(\omega,n)}=0\,.
\end{equation}
The later equation has $2\,d$ roots $\omega=\omega^{(n)}_J$, $J=1,\dots,2\,d$, which are the frequencies from (\ref{freq base ans}). The ansatz (\ref{freq base ans}) is justified only if all the roots are distinct, or if equal roots are associated
with linearly independent null eigenvectors,
because it involves all the $2\,d$ linearly independent solutions of the equation (\ref{eom transv}). The degenerate case requires separate considerations. In what follows we will always assume distinct roots.
From (\ref{The M matrix}) one immediately notes that $M^T(\omega,n)=M(-\omega,-n)$. This property means that $M(\omega,n)$ and $M(-\omega,-n)$ have the same determinant and hence the same roots. This leads to the situation where for $n=0$ the frequencies come in pairs, $\{\omega_J\}=\{\pm\omega_j, \,j=1,\dots,d\}$. It is then convenient to rewrite
the expansion of the zero-mode as
\begin{equation}
X_0^i(\tau)=\sum^d_{j=1}\left(\zeta^{+}_j\,a^{+}_{ij} \,e^{i\,\omega_j\,\tau}+\zeta^{-}_j\,a^{-}_{ij} \,e^{-i\,\omega_j\,\tau}\right)\,.
\end{equation}
For the higher modes ($n\neq 0$) the $\pm n-$modes are paired, $\omega^{(n)}_J=-\omega^{(n)}_j$, $J=1,\dots,2\,d$. It is useful to chose the eigen-directions $a^{(n)}_{iJ}$ in the following way:
\begin{equation}
a^{(n)}_{iJ}=(-1)^i\,m_{1i}\left(\omega^{(n)}_J\right)\,,
\end{equation}
where $m_{ij}\left(\omega^{(n)}_J\right)$, $i,\,j=1,\dots,d$, are the minors $m_{ij}$ of the matrix $M(\omega,n)$, evaluated for $\omega=\omega_J^{(n)}$. Therefore one can rewrite the solution for the string modes explicitly as
\begin{equation}
X^i_0(\tau)=(-1)^i\,\sum^d_{j=1}\left(\zeta^{+}_j\,m_{1i}(\omega_j) \,e^{i\,\omega_j\,\tau}+\zeta^{-}_j\,m_{i1}(\omega_j) \,e^{-i\,\omega_j\,\tau}\right),\quad n=0\,,
\end{equation}
\begin{equation}
X^i_n(\tau)=(-1)^i\,\sum^{2\,d}_{J=1}\zeta^{(n)}_J\,m_{1i}(\omega_J^{(n)}) \,e^{i\,\omega_J^{(n)}\,\tau}\,,\quad n\neq 0\,.
\end{equation}
In order to find the Hamiltonian we promote the $\zeta$’s to operators with  commutation relations given by
\begin{equation}
C_j=[\zeta^{-}_j,\zeta^{+}_j]\,,\quad C^{(n)}_J=[\zeta^{(-n)}_J,\zeta^{{n}}_J]\,,\quad C^{(-n)}_J=-C^{(n)}_J\,,
\end{equation}
where one has
\begin{equation}
C_j=\frac{1}{2\,m_{11}(\omega_j)\,\omega_j\prod_{k\neq j}(\omega^2_j-\omega_k^2)}\,,\qquad
C_J^{(n)}=\frac{1}{m_{11}(\omega_J^{(n)})\,\prod_{K\neq J}(\omega^{(n)}_J-\omega_K^{(n)})}\,.
\end{equation}
The expressions for these coefficients follow from the canonical equal-time commutation relations between the string modes $X$. The relations between the $\zeta$'s and the canonically normalised operators $a_j^{\pm}$, $[a^{-}_j,a^{+}_k]=\delta_{jk}$, are given by
\begin{equation}
a^{\pm\sigma}_j=\frac{\zeta^{\pm}_j}{\sqrt{|C_j|}}\,,
\end{equation}
where $\sigma={\rm{sign}}(C_j)$. With this choice for the $a$'s the bilinear combination $\zeta^{+}_j\,\zeta^{-}_j+\zeta^{-}_j\,\zeta^{+}_j$ is related to the number operator, $\mathcal{N}_j=a^{+}_j\,a^{-}_j$, by
\begin{equation}
\frac{1}{2}\,(\zeta^{+}_j\,\zeta^{-}_j+\zeta^{-}_j\,\zeta^{+}_j)=|C_j|\,\left(\mathcal{N}_j+\frac{1}{2}\right)\,.
\end{equation}
Similar relation holds for the higher modes,
\begin{equation}
\frac{1}{2}\,\left(\zeta^{(n)}_J\,\zeta^{(-n)}_J+\zeta^{(-n)}_J\,\zeta^{(n)}_J\right)=|C_J^{(n)}|\,\left(\mathcal{N}_J^{(n)}+\frac{1}{2}\right)\,,
\end{equation}
which connects the number operator $\mathcal{N}_J^{(n)}=a^{(n)}_J\,a^{(-n)}_J$ with the $\zeta$ operators.
Hence, one has an explicit string mode expansion. Thus, the string Hamiltonian,
\begin{equation}
H=\frac{1}{2\,\pi}\,\int^\pi_0d\sigma \left[\delta_{ij}\left(\dot{X}^i\,\dot{X}^j+X^{i\,\prime}\,X^{j\,\prime}-k_i\,X^i\,X^j\right)-2 \,h_ij\,X^i\,X^{j\,\prime}\right]\,,
\end{equation}
can be written as a sum of $n$-level harmonic oscillator Hamiltonians
\begin{equation}
H=\sum^{\infty}_{n=0} H^{(n)}\,.
\end{equation}
Here the zero-mode part Hamiltonian assumes the form
\begin{equation}\label{zeromode H}
H^{0}=\sum^{d}_{j=1}\mathrm{sign}\left(C_j\right)\,\Omega_j\,\left(\mathcal{N}_j+\frac{1}{2}\right)\,,
\end{equation}
with frequencies
\begin{equation}
\Omega_j= \frac{\sum_{i=1}^{d}(\omega_j^2-k_i)\,m_{ii}(\omega_j)}{2\,\omega_j\prod_{k\neq j}(\omega^2_j-\omega_k^2)}\,.
\end{equation}
Likewise, the Hamiltonians for higher modes of the string are given by
\begin{equation}
H^{(n)}=\sum^{2\,d}_{J=1}\mathrm{sign}\left(C_J^{(n)}\right)\,\Omega_J^{(n)}\,\left(\mathcal{N}_J^{(n)}+\frac{1}{2}\right)\,,\quad n>0\,,
\end{equation}
where the frequency $\Omega^{(n)}_J$ is a sum of two terms -- one coming from the plane wave metric, and the other coming from the Kalb-Ramond B-field:
\begin{equation}
\Omega^{(n)}_J=2\,\omega^{(n)}_J\,C^{(n)}_J\,m_{11}\left(\omega^{(n)}_J\right)\,\sum_{i,j}\left(\omega^{(n)}_J\,\delta_{ij}+i\,(-1)^{i+j}\,f_{ij}\right)\,m_{ij}\left(\omega^{(n)}_J\right)\,.
\end{equation}
\subsection{Extended entanglement entropy in the ground state of the bosonic string }
\begin{figure}[H]
\begin{center}
\includegraphics[scale=0.7]{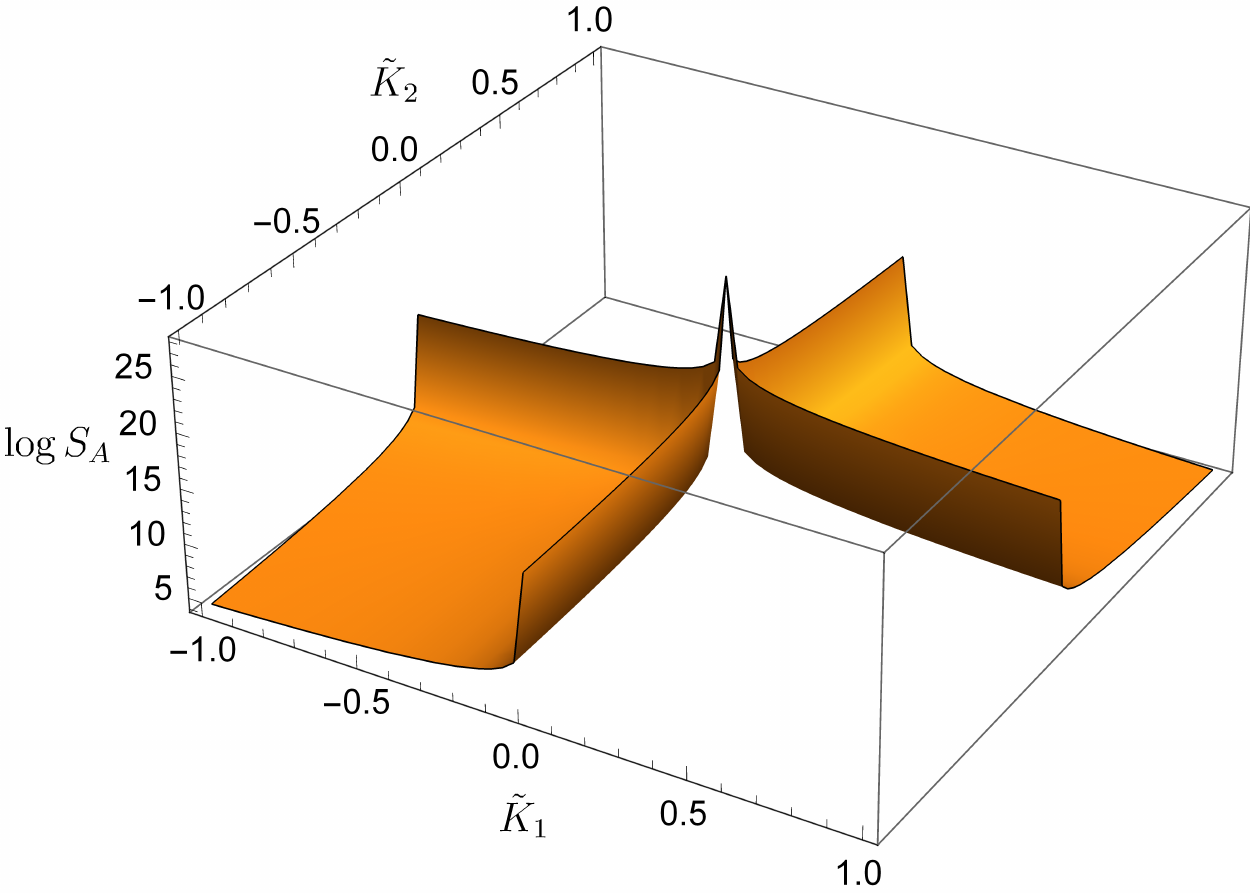}
\caption{The renormalized entanglement entropy for closed string in 5-dimensional regular plane wave background, $k_B=1$. At the origin (at very high temperatures) EREE diverges.\label{figEREE5d}}
\end{center}
\end{figure}

We are now ready to apply the TFD technique for the entanglement entropy on every energy level of the string spectrum. Here, for convenience, we consider only the $n=0$ Hamiltonian of the string from eq. (\ref{zeromode H}). Assume the following two subsystems:
\begin{equation}
\left\{ {{{\cal N}_j}} \right\}_{j = 1}^d = \left\{ {{{\cal N}_\mu}} \right\}_{\mu  = 1}^p\bigcup {\left\{ {{\mathcal{N}_k}} \right\}_{k = p + 1}^d} \,,\quad p \le d - 1,\quad 2 \le d \le 9\,,
\end{equation}
the resulting entanglement entropy agrees with equation (\ref{EE general}):
\begin{equation}
{S_A}\left( {{{\tilde K}_\mu }} \right) = \frac{{{1}}}{2}\,\left( {\prod\limits_{\mu  = 1}^p {\coth \frac{{{{\tilde K}_\mu }}}{4}} } \right){\mkern 1mu} \,\sum\limits_{\mu  = 1}^p {\left\{ {{K_\mu }{\mkern 1mu} \left( {1 + \coth \frac{{{{\tilde K}_\mu }}}{4}} \right) - 2\,\ln \left( {{e^{{{\tilde K}_\mu }}} - 1} \right)} \right\}} \,.
\end{equation}
The thermal parameters, ${{\tilde K}_\mu } = \beta \,{\rm{sign}}\,\left( {{C_\mu }} \right)\,{\Omega _\mu }$, depend on the frequencies of the classical string modes $\omega_\mu$, $\mu=1,\dots,p$. In four-dimensional spacetime $d=2$, thus $p=1$, the space of parameters is one-dimensional spanned by the values of $\tilde K_1$. The entanglement entropy behaves as shown in figure \ref{figEREEtoEE}. In 5-dimensions ($d=3$, $p=2$), the space of parameters is a two-dimensional Riemannian manifold spanned by $(\tilde K_1,\,\tilde K_2)$. The the entanglement entropy is given by
\begin{align}\label{EE strings}
\nonumber &{S_A}({\tilde K_1},{\tilde K_2}) = \frac{{1}}{2}\,\coth \frac{{{\tilde K_1}}}{4}\,\coth \frac{{{\tilde K_2}}}{4}
\\
&\times\left[ {{\tilde K_1}\,\left( {1 + \coth \frac{{{\tilde K_1}}}{4}} \right) + {\tilde K_2}\,\left( {1 + \coth \frac{{{\tilde K_2}}}{4}} \right)} \right.\left. { - 2\,\log \left[ {\left( {{e^{{\tilde K_1}}} - 1} \right)\,\left( {{e^{{\tilde K_2}}} - 1} \right)} \right]\,\coth \frac{{{\tilde K_1}}}{4}} \right]\,.
\end{align}
The inverse scaled temperatures, $\tilde K_{1,2}$, depend on the sign of the coefficients $C_{1,2}$, which leads to two regions on the plot (fig. \ref{figEREE5d}) -- one for positive values of the $K$'s, and one for negative ones.
\\
\noindent As expected for very high temperatures ($\tilde K_i\to 0$) the entropy diverges. One finds similar situation for the standard entanglement entropy,
\begin{equation}
{\Sigma _A}({\tilde K_1},{\tilde K_2}) = \frac{{4\,{\tilde K_1}}}{{{e^{{\tilde K_1}}} - 1}} + \frac{{4\,{\tilde K_2}}}{{{e^{{\tilde K_2}}} - 1}} - \log \left[ {\left( {1 - {e^{ - {\tilde K_1}}}} \right)\,\left( {1 - {e^{ - {\tilde K_2}}}} \right)} \right]\,,
\end{equation}
shown in figure \ref{figNEE} below,
\begin{figure}[H]
\begin{center}
\includegraphics[scale=0.7]{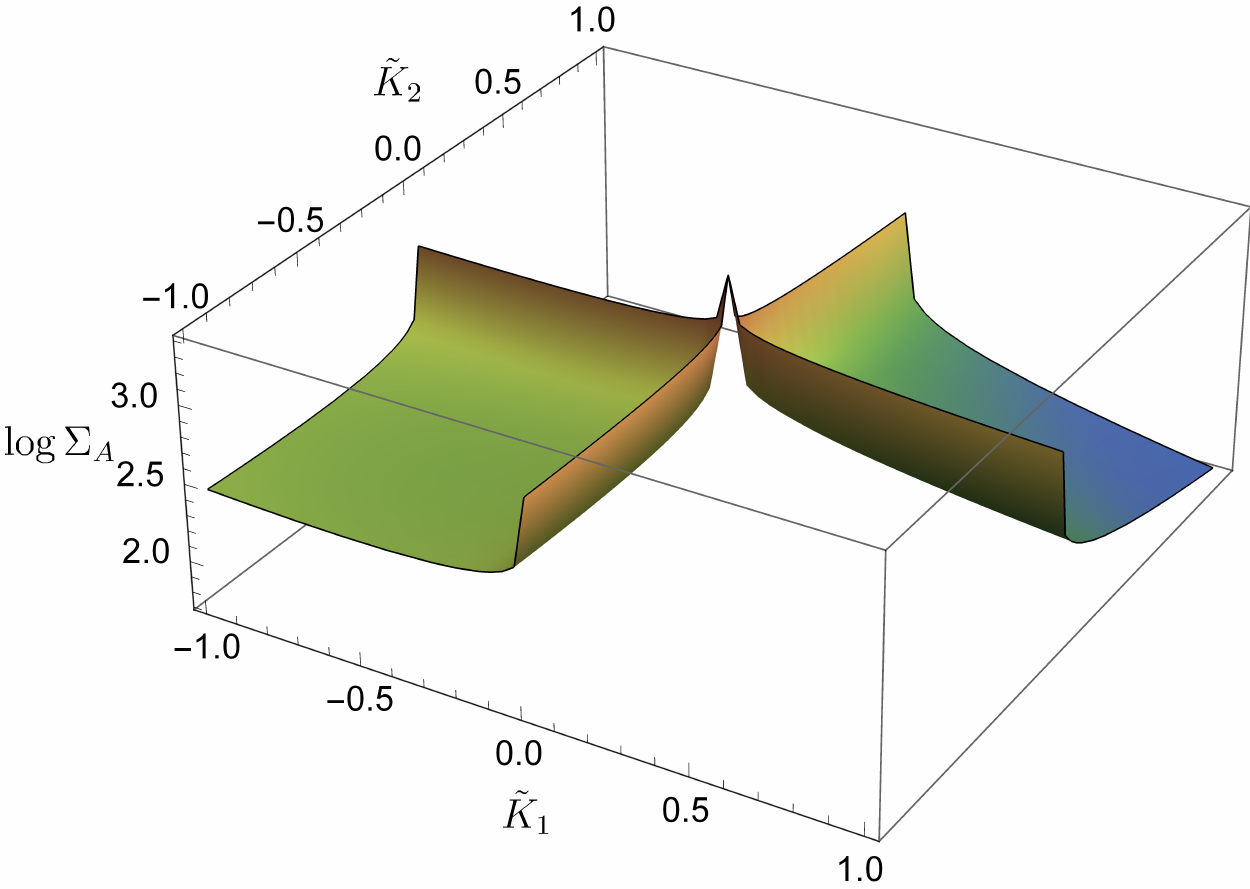}
\caption{The standard entanglement entropy for closed bosonic string in the 5-dimensional regular plane wave background.\label{figNEE}}
\end{center}
\end{figure}
The Fisher metric, $g_{\mu\nu}=\partial_\mu\partial_\nu S$, at the point ($\tilde K_1=0,\tilde K_2=0$), is also singular, as can be seen from the following expressions for the metric coefficients:
\begin{align}\label{2dfisher}
	g_{11}&=\frac{1}{64}\,\coth\frac{\tilde K_2}{4}\csch^2\frac{\tilde K_1}{4}\left[\tilde K_1\left(3+5\coth^2\frac{\tilde K_1}{4}
	+7\csch^2\frac{\tilde K_1}{4}\right)+4\tanh\frac{\tilde K_1}{4}\right.\nonumber\\
	&\phantom{{}={}}\left.+4\coth\frac{\tilde K_1}{4}\left(\tilde K_1+\tilde K_2-5+\tilde K_2\coth\frac{\tilde K_2}{4}
	-2\log\left[\left(e^{\tilde K_1}-1\right)\left(e^{\tilde K_2}-1\right)\right]\right)\right]\,,\\
	\label{2dfisher2}
	g_{12}&=g_{21}=\frac{1}{32}\,\csch^2\frac{\tilde K_1}{4}\csch^2\frac{\tilde K_2}{4}
	\left[\tilde K_1\left(1+2\coth\frac{\tilde K_1}{4}\right)+\tilde K_2\left(1+2\coth\frac{\tilde K_2}{4}\right)-4\right.\nonumber\\
	&\phantom{{}={}}\left.-2\log\left[\left(e^{\tilde K_1}-1\right)\left(e^{\tilde K_2}-1\right)\right]\vphantom{\coth\frac{\tilde K_2}{4}}\right]\,,\\
	\label{2dfisher3}
	g_{22}&=\frac{1}{64}\,\coth\frac{\tilde K_1}{4}\csch^2\frac{\tilde K_2}{4}\left[\tilde K_2\left(3+5\coth^2\frac{\tilde K_2}{4}
	+7\csch^2\frac{\tilde K_2}{4}\right)+4\tanh\frac{\tilde K_2}{4}\right.\nonumber\\
	&\phantom{{}={}}\left.+4\coth\frac{\tilde K_2}{4}\left(\tilde K_1+\tilde K_2-5+\tilde K_1\coth\frac{\tilde K_1}{4}
	-2\log\left[\left(e^{\tilde K_1}-1\right)\left(e^{\tilde K_2}-1\right)\right]\right)\right]
\end{align}
This result is already familiar \cite{DMRV:2016}. The singular point at the origin is a signal of a phase transition. As shown from the geometric analysis of the Fisher metric in section \ref{sec GAFMPT} its Ricci scalar is regular at the origin, which suggests that the point ($K_1=0,K_2=0$) is not a second order phase transition. Furthermore, the scalar curvature at that point is zero, corresponding to a free quasi-system at very high temperatures.
\subsection{Geometric analysis of the Fisher metric and phase transitions}\label{sec GAFMPT}

\begin{figure}[H]
\begin{center}
\includegraphics[scale=0.7]{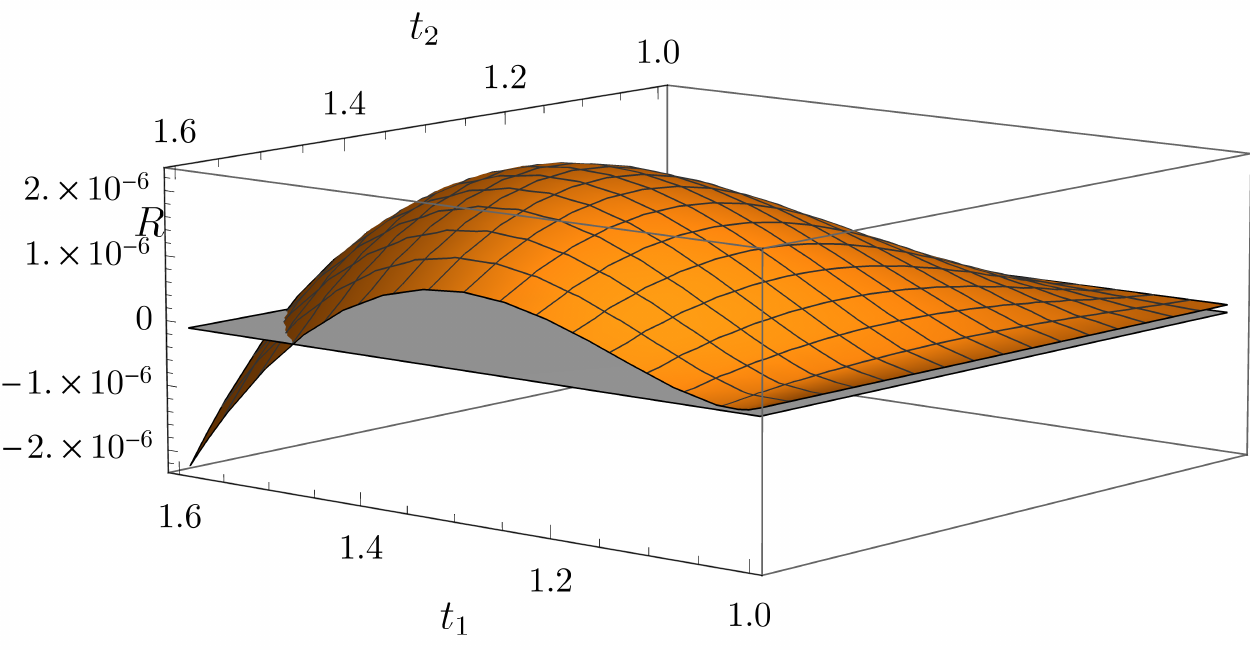}
\caption{Visualization of the scalar curvature $R$ in terms of $t_1,\,t_2$, near the origin ($t_1=1,\,t_2=1$). The curvature is positive defined implying elliptic geometry on the statistical manifold of thermodynamic parameters. The local maximum corresponds to the strongest interaction in the quasi-system.\label{curvature t}}
\end{center}
\end{figure}

\begin{figure}[H]
\begin{center}
\includegraphics[scale=0.6]{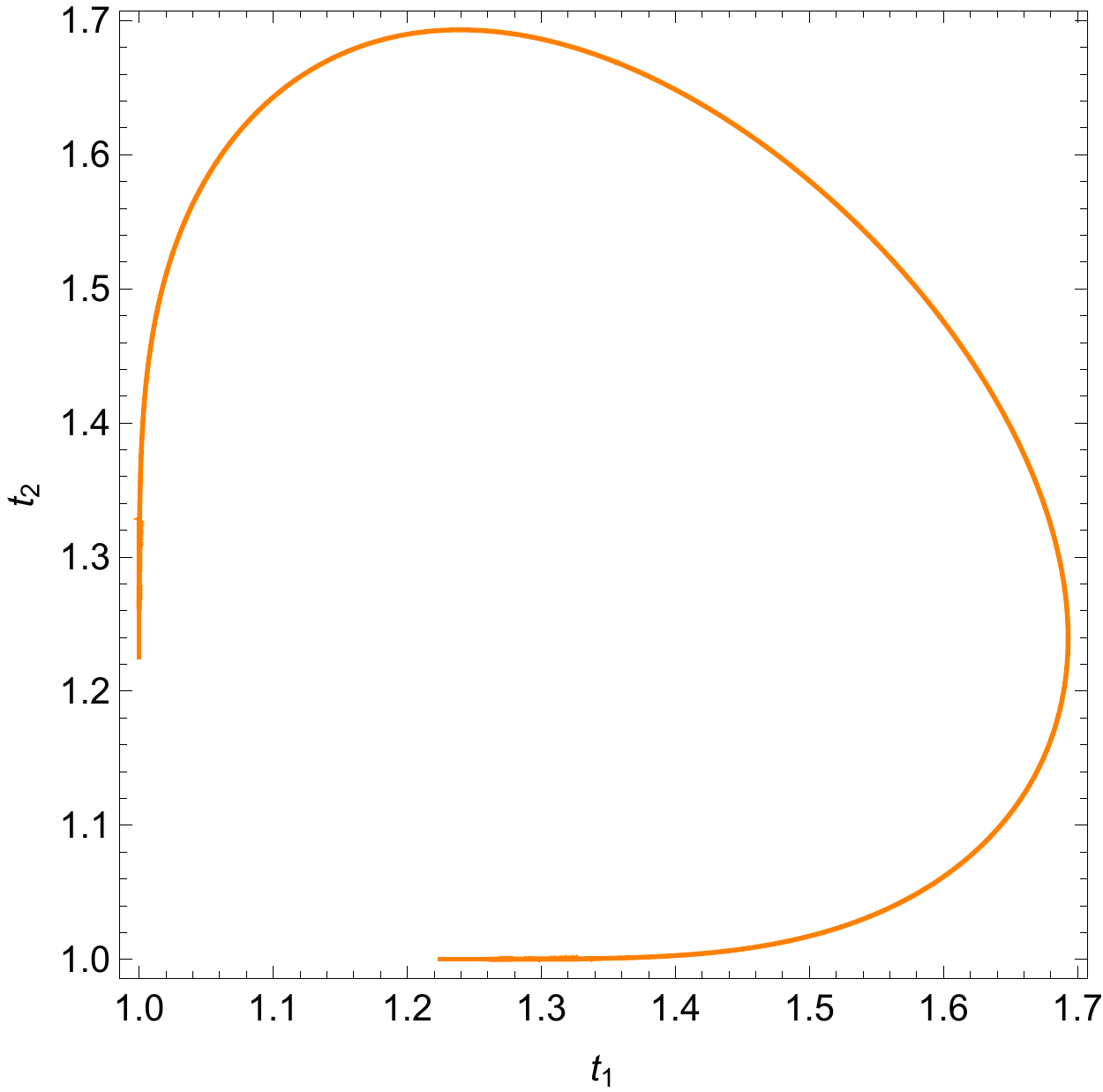}
\caption{The contour plot of the level curve $R=0$, corresponding to the values of the parameters for which the quasi-system is effectively free. In this picture one should also include the origin (high-temperatures), $t_1=t_2=1$, and infinity (low-temperatures), $t_1,\,t_2\to\infty$.\label{curvature zero} }
\end{center}
\end{figure}

Well-known fact is that the Fisher information metric defines a Riemannian metric on the space of parameters \cite{BR:1984, AN:2000, AM:2016} for variety of statistical systems. Such geometrization is often useful in the analysis of the phase structure for a given statistical model \cite{BR:1999, GR:1995}. Here the scalar curvature, $R$, plays a central role, e.g. a non-interacting model shows a flat geometry ($R = 0$), while $R$ diverges at the critical points of an interacting one, thus effectively preventing
geodesics from crossing into the nonphysical area of phase space \cite{BR:2003, BR:2009, KMS:2012, MMS:2015}. The specific critical points, where the phase transition occurs, lie on the spinodal curve. An advantage of the probabilistic description of the system's phase structure is that one does not require the definition of order parameters. This is useful for systems where an order parameter is difficult to identify, or does not exist.
\\
\indent In what follows we analyse the scalar curvature $R$ of the Fisher metric from (\ref{2dfisher}), (\ref{2dfisher2}) and (\ref{2dfisher3}). The scalar curvature is independent of the chosen coordinates, so, for convenience, we perform a change of variables from $K_1$ and $K_2$ to $t_1=e^{K_1}$ and $t_2=e^{K_2}$. The Fisher metric in the new coordinates is given by
\small
\begin{align}
\nonumber &{g_{11}} = \frac{1}{{4\,t_1^{3/2}\,{{(T_1^ - )}^4}\,T_1^ + \,{{\left( {T_2^ - } \right)}^2}}}\,\left\{ {T_1^ + \,\left( {1 + 8\,\sqrt {{t_1}}  + 3\,{t_1}} \right)\,\left( {{t_2} - 1} \right)\,\log {t_1}} \right.\\
 &+ 2\,T_1^ - \,\left. {\left( {2\,\left( {1 + 3\,\sqrt {{t_1}}  + {t_1}} \right)\,\left( {1 - {t_2}} \right) + {{\left( {T_1^ + } \right)}^2}\,\left( {\left( {1 - {t_2}} \right)\,\log \left[ {\left( {{t_1} - 1} \right)\,\left( {{t_2} - 1} \right)} \right] + \left( {\sqrt {{t_2}}  + {t_2}} \right)\,\log {t_2}} \right)} \right)} \right\}\,,\\
 \nonumber
&{g_{12}} = {g_{21}} = \frac{1}{{2\,{t_2}\,\sqrt {{t_1}} \,{{\left( {T_1^ - } \right)}^3}\,{{\left( {T_2^ - } \right)}^3}}}\,\left\{ {\left( {{t_2} - \sqrt {{t_2}} } \right) \times } \right.\\
 &\times \left( {\left( {1 + 3\,\sqrt {{t_1}} } \right)\,\log {t_1} - 2\,T_1^ - \,\left( {2 + \log \left[ {\left( {{t_1} - 1} \right)\,\left( {{t_2} - 1} \right)} \right]} \right)} \right) + \left. {T_1^ - \left( {\sqrt {{t_2}}  + 3\,{t_2}} \right)\,\log {t_2}} \right\}\,,\\\nonumber\\\nonumber
&{g_{22}} = \frac{1}{{4\,t_2^{3/2}\,T_2^ + \,{{\left( {T_1^ - } \right)}^2}\,{{\left( {T_2^ - } \right)}^4}}}\,\left\{ {2\,T_1^ + \,T_2^ - \,\sqrt {{t_1}} \,{{\left( {T_2^ + } \right)}^2}\,\log {t_1}} \right.\\
 &+ \left( {{t_1} - 1} \right)\,\left. {\left( { - 2\,T_2^ - \,\left( {2\,\left( {1 + 3\,\sqrt {{t_2}}  + {t_2}} \right) + {{\left( {T_2^ + } \right)}^2}\,\log \left[ {\left( {{t_1} - 1} \right)\,\left( {{t_2} - 1} \right)} \right]} \right) + T_2^ + \,\left( {1 + 8\,\sqrt {{t_2}}  + 3\,{t_2}} \right)\,\log {t_2}} \right)} \right\}\,,
\end{align}
where $1\leq t_1,\,t_2\leq\infty$ and
\begin{equation}
T_1^ -  = \sqrt {{t_1}}  - 1\,,\quad T_1^ +  = 1 + \sqrt {{t_1}} \,,\quad T_2^ -  = \sqrt {{t_2}}  - 1\,,\quad T_2^ +  = 1 + \sqrt {{t_2}} \,.
\end{equation}
\normalsize
The explicit expression for $R$ is too lengthy to be presented here. However its functional dependence on ($t_1,\,t_2$) near the origin (1,1) is shown on figure \ref{curvature t}. One notes that the Ricci scalar is positive defined and shows local maximum near the point ($t_1=1.3, \,t_2=1.3$). The positive values of the scalar curvature suggest elliptic geometry in the thermodynamic parameter space, while the local maximum corresponds to the strongest interaction between the constituents of the quasi-system. There is a level curve $k(t_1,t_2)$ for which $R=0$, corresponding to free non-interacting system (fig. \ref{curvature zero}). At the origin the scalar curvature is also regular and tends to zero, which implies that the singular point $(K_1=0,K_2=0)$ in the Fisher metric is not a second-order phase transition and also shows that at very high temperatures the system is free. One notes that the values of $R$ do not deviate much from zero, which makes the entire quasi-system almost non-interacting. This kind of behaviour is expected due to the properties of the Bogoliubov transformation, which smoothens out the strength of the interactions in the original quantum system and effectively produces a non-interacting quasi-system.

\begin{figure}[H]
\begin{center}
\includegraphics[scale=0.7]{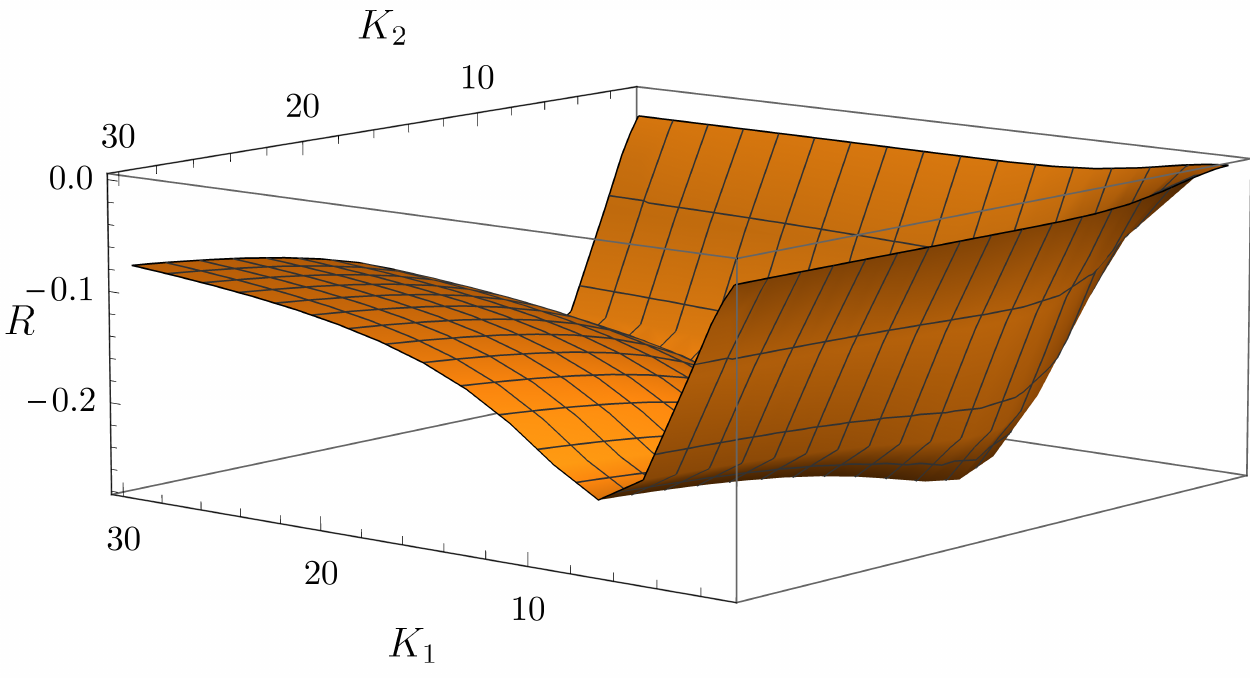}
\caption{The behaviour of the scalar curvature $R$ in terms of $K_1,\,K_2$ for large values of the inverse scale temperatures. There is an obvious local minimum, corresponding to the strongest interaction in the hyperbolic case.\label{curvature k}}
\end{center}
\end{figure}

\noindent For larger values of the parameters the Ricci scalar is not positive defined as shown in figure \ref{curvature k}. In this case the geometry in the space of parameters is hyperbolic. The non-zero values of the scalar curvature suggest also interacting system. There is a local minimum, corresponding to the highest strength of the interactions in the hyperbolic case. For low temperatures ($K_{1,2}\to\infty$) the scalar curvature tends to zero once again corresponding to free non-interacting quasi-system.
The TFD scalar curvature is regular for all points from the two-dimensional space of parameters. Therefore one concludes that the closed bosonic string system in 5-dimensional homogeneous plane wave background does not show any second-order phase transitions.

\section{Non-equilibrium entanglement entropy for dissipative systems}\label{sec NEEEDS}
\subsection{Extended entanglement entropy for dissipative system}
\begin{figure}[H]
\begin{center}
\includegraphics[scale=1.1]{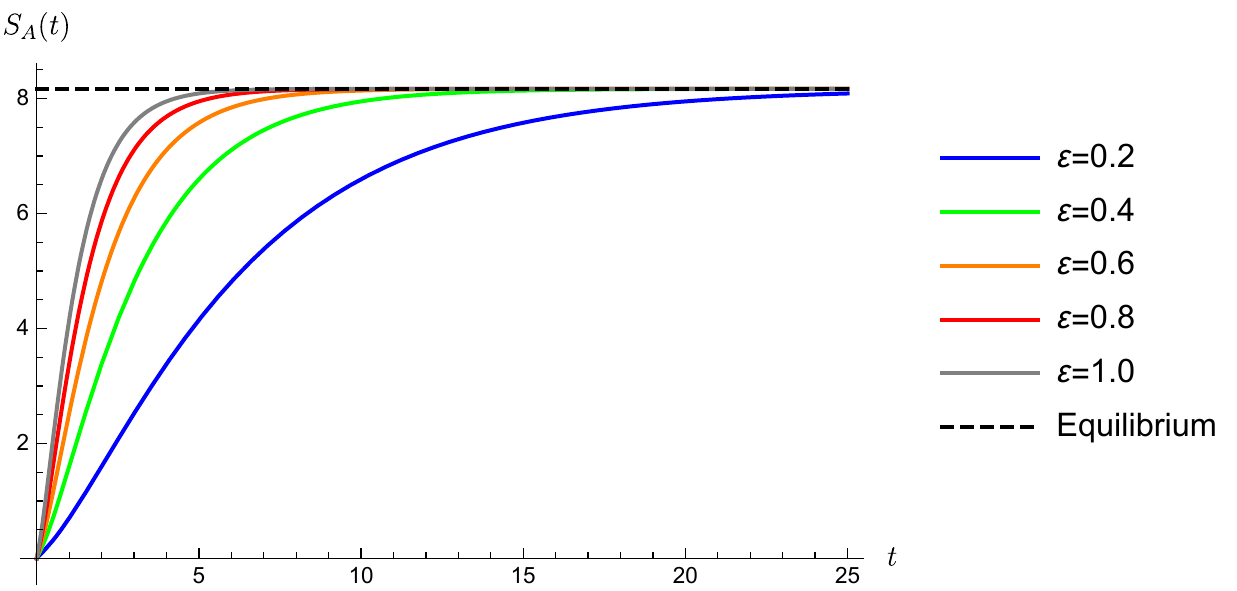}
\caption{Non-equilibrium EREE as a function of time in the $p=1$ case (also $k_B=1$ and $K_1=1$). The dashed line is the equilibrium value of the entropy saturated at $t\to\infty$. Different curves correspond to different values of $\varepsilon$ in the interval [0,1].\label{figEREEofT}}
\end{center}
\end{figure}

\begin{figure}[H]
\begin{center}
\includegraphics[scale=1.1]{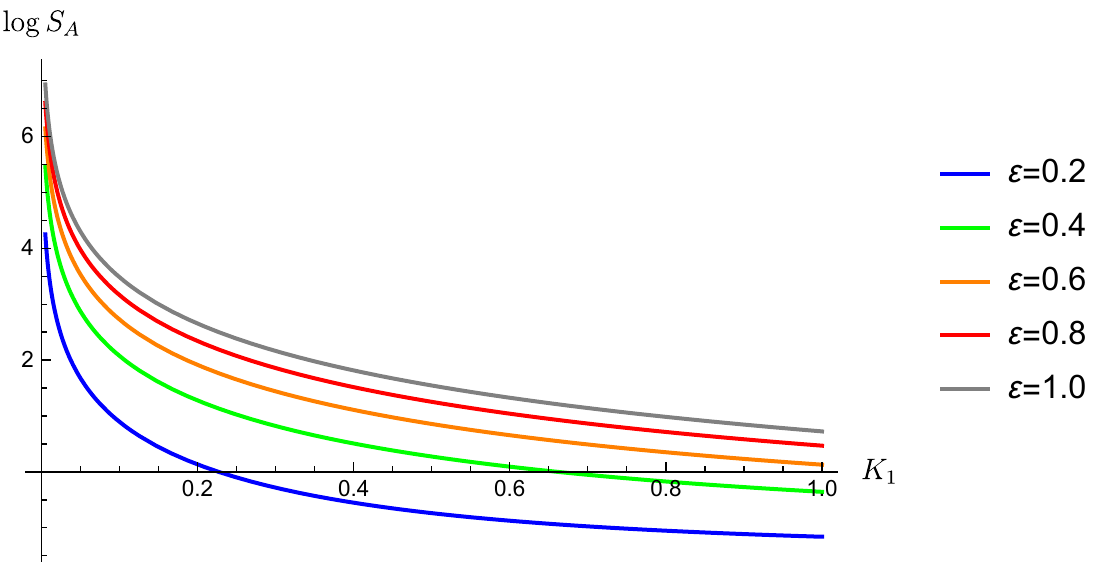}
\caption{Non-equilibrium EREE ($p=1$) as a function of the scaled inverse temperature parameter $K_1$ at fixed moment $t=0.5$, for different values of the dissipation parameter $\varepsilon\in[0,1]$.\label{figEREEofK}}
\end{center}
\end{figure}

Following \cite{Nakagawa:2016}, one can consider the Hamiltonian from eq. (\ref{HamilMatrixStatic}) as a non-equilibrium system with dissipations. In this case the time-dependent density operator $\hat \rho_{neq}(t)$ satisfies the dissipative von Neumann equation:
\begin{equation}\label{}
i\,{\partial _t}\hat \rho_{neq} (t) = \left[ {\hat H,\,\hat \rho_{neq} (t)} \right] - \varepsilon \,\left( {\hat \rho_{neq} (t) - {{\hat \rho }_{eq}}} \right)\,,
\end{equation}
where $\varepsilon$ is a dissipation parameter and $\hat \rho_{eq}$ is defined in eq. (\ref{equilDensityMatrix}).
The solution to this equation is formally given by
\begin{equation}\label{solDensityOperatorNonequilCase}
\hat \rho_{neq} (t) = {e^{ - \varepsilon \,t}}\,{{\hat U}^\dag }(t)\,{{\hat \rho }^{(0)}}\,\hat U(t) + (1 - {e^{ - \varepsilon \,t}})\,{{\hat \rho }_{eq}}\,,
\end{equation}
where $\hat\rho^{(0)}$ is an arbitrary initial density matrix and $\hat U(t):=e^{i\,t\,\hat H}$. The diagonal form of the Hamiltonian allows one to write
\begin{equation}\label{EvolutorDiagonal}
\hat U(t) = \sum\limits_{\left\{ {{n_i}} \right\} = 0}^\infty  {{e^{i\,t\,\left( {\sum\limits_{i = 1}^N {{E_i}\,{n_i} + {E_0}} } \right)}}} \,\left| {\left\{ {{n_i}} \right\}} \right\rangle \left\langle {\left\{ {{n_i}} \right\}} \right|\,.
\end{equation}
The case of arbitrary initial conditions significantly complicates the calculations. Therefore one can consider only initial conditions in the ground state as suggested in ref. \cite{Nakagawa:2016}:
\begin{equation}
\hat \rho^{(0)}=\frac{e^{-K_0}}{Z}\,|\{0\}\rangle\langle \{0\}|\,.
\end{equation}
Here ${K_0} = \beta \,{E_0}$, $E_0$ is the energy in the ground state, and $Z(K_i)$ is given by eq. (\ref{statSumEqCase}). Once again we use the simple bipartition of the bulk system, namely
\begin{equation}
\left\{ {{n_i}} \right\}_{i = 1}^N = \left\{ {{n_A}} \right\}_{A = 1}^p\bigcup {\left\{ {{n_B}} \right\}_{B = p + 1}^N} \,,\quad p \le N - 1,\quad N \ge 2\,.
\end{equation}
Following the steps shown in \cite{Hashizume:2013}, one finds the renormalized extended entanglement entropy in the form ($k_B=1$):
\begin{align}\label{general EE non-eq case}
{S_A}({K_i},\varepsilon ;t) =  - {e^{ - \varepsilon \,t}}\,a(t)\,\log a(t) + (1 - {e^{ - \varepsilon \,t}})\,\sum\limits_{\mu  = 1}^p {\left( {{S_\mu (t)}\,\tanh
 \frac{{{K_\mu }}}{4}} \right)} \,\prod\limits_{\alpha  = 1}^p {\coth \frac{{{K_\alpha }}}{4}} \,,
\end{align}
where $i=1,\dots,p,p+1,\dots,N$, and
\begin{align}
\nonumber {S_\mu (t)} &=  - {a_\mu }(t)\,\log {a_\mu }(t) - \frac{1}{2}\,{e^{ - {K_\mu }}}\,\coth \frac{{{K_\mu }}}{4}\\
 \nonumber &\times \left\{ {b(t)\,\left( {{e^{{K_\mu }/2}} - 1} \right){\,} \left[ {4\,\log b(t) + 4\,\log \left( {{e^{{K_\mu }}} - 1} \right) - {K_\mu }\,\left( {5 + \coth \frac{{{K_\mu }}}{4}} \right)} \right]} \right.\\
& + \left. {(1 - {e^{ - \varepsilon \,t}})\,\left[ {2\,\log \left( {1 - {e^{ - \varepsilon \,t}}} \right) + 2\,\log \left( {{e^{{K_\mu }}} - 1} \right) - {K_\mu }\,\left( {3 + \coth \frac{{{K_\mu }}}{4}} \right)} \right]} \right\}\,,
\end{align}
\begin{equation}
a(t) = \left( {(1 - {e^{ - \varepsilon {\kern 1pt} t}})\,\prod\limits_{r = p + 1}^N {{{\left( {1 - {e^{ - {K_r}}}} \right)}^{ - 1}}}  + {e^{ - \varepsilon \,t}}} \right){\mkern 1mu} \,\prod\limits_{\mu = 1}^p {\left( {1 - {e^{ - {K_\mu }}}} \right)\,} \prod\limits_{r = p + 1}^N {\left( {1 - {e^{ - {K_r}}}} \right)} \,,
\end{equation}
\begin{equation}
b(t) = \left( {\sqrt {1 - {e^{ - \varepsilon \,t}}}  - (1 - {e^{ - \varepsilon \,t}})\left( {\prod\limits_{r = p + 1}^N {{{\left( {1 - {e^{ - {K_r}}}} \right)}^{ - 1}}}  - 1} \right)} \right)\,\prod\limits_{r = p + 1}^N {\left( {1 - {e^{ - {K_r}}}} \right)} \,,
\end{equation}
\begin{equation}
{a_\mu }(t) = \left( {1 - {e^{ - {K_\mu }}}} \right)\,\left( {(1 - {e^{ - \varepsilon \,t}})\,\prod\limits_{r = p + 1}^N {{{\left( {1 - {e^{ - {K_r}}}} \right)}^{ - 1}}}  + {e^{ - \varepsilon \,t}}} \right)\,\prod\limits_{r = p + 1}^N {\left( {1 - {e^{ - {K_r}}}} \right)} \,.
\end{equation}
The limit, $t\to \infty$, coincides with the equilibrium case (\ref{EE generalExp}):
\begin{equation}
\mathop {\lim }\limits_{t\to \infty } {S_A}({K_i},\varepsilon ;t) = \frac{1}{2}\,\left( {\prod\limits_{\mu  = 1}^p {\coth \frac{{{K_\mu }}}{4}} } \right){\mkern 1mu} \,\sum\limits_{\mu  = 1}^p {\left\{ {{K_\mu }\,\left( {1 + \coth \frac{{{K_\mu }}}{4}} \right) - 2\,\ln \left( {{e^{{K_\mu }}} - 1} \right)} \right\}} \,.
\end{equation}
The limit at $t\to 0$ reduces to the ground state:
\begin{equation}
\lim_{t\to0}{S_A}\left( {{K_\mu },\varepsilon ;t} \right) = -\frac{e^{-K_0}}{Z} \,\log{\frac{e^{-K_0}}{Z}}.
\end{equation}
On figure \ref{figEREEofT} is shown the time dependence of the entropy for several values of the dissipation parameter $\varepsilon$. Clearly, given enough time, the entropy reaches the equilibrium case from (\ref{EE general}) (depicted as a black dashed line). This is consistent with the second law of thermodynamics. 
\indent One also depicts a generic behaviour of the entropy, which has been observed for various systems prepared in a state with initially
low entanglement entropy, $S_A(t_0)\leq S_{eq}$: after some short transient period of time, which depends on
the initial state of the system, the entanglement entropy goes through a phase of linear or almost linear growth, ${S_A}(t)\sim{\Lambda _A}\,t$, until it settles at the saturation phase. This kind of behaviour is observed
in the time evolution of various quantum systems that bear the signatures of quantum chaos \cite{Zurek:1994wd, 1999chao.dyn.11017P, 2000PhRvL..85.3373M, 2002PhRvE..66d5201T}, or in the study of thermalization in some holographic systems \cite{Balasubramanian:2010ce, Balasubramanian:2011ur, Hartman:2013qma, Liu:2013qca}. In particular,
the rate of growth $\Lambda_A$ of the entanglement entropy
in the phase of linear growth is connected to the scrambling time in chaotic quantum systems, or the Kolmogorov-Sinai entropy rate \cite{Bianchi:2017kgb}, in cases where the quantum system has a classical counterpart. 
\\
\indent Finally, on figure \ref{figEREEofK} we show the logarithm of the entanglement entropy as a function of the scaled inverse temperature parameter $K_1$ at fixed finite moment of time $t$. As expected the entropy decreases with increasing $K_1$.
\subsection{Entanglement entropy production rate}
\begin{figure}[H]
\begin{center}
\includegraphics[scale=1.1]{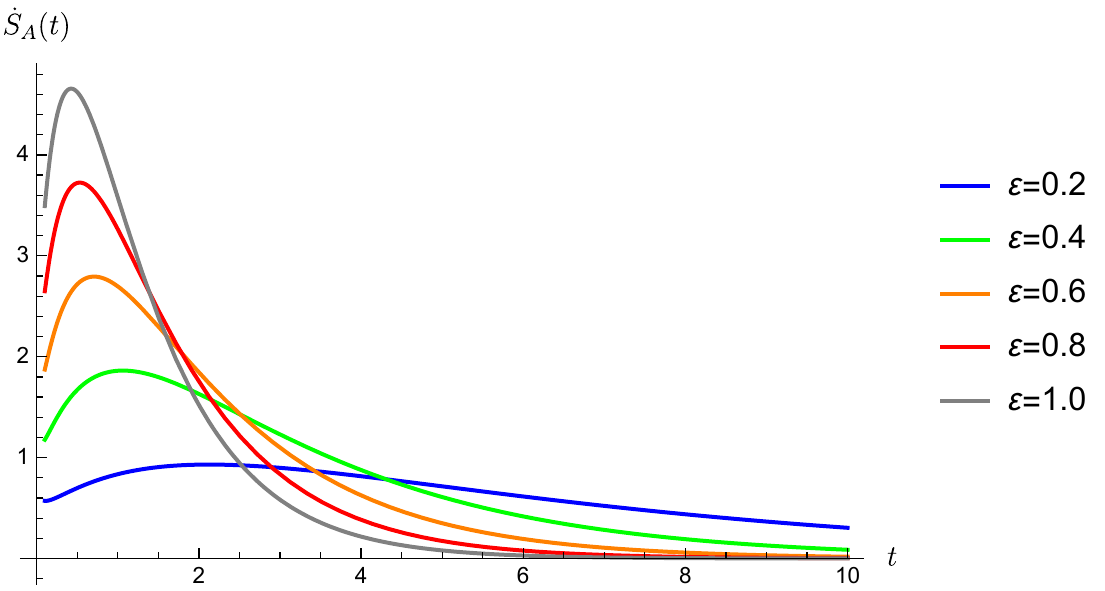}
\caption{The entropy increase rate $\dot{S}_A(t)$ for several values of the dissipation parameter $\varepsilon$. One notes that for short time the entropy production rate reaches peak value, after which it monotonously decreases with time. Also the peak entropy production is bigger for strongly dissipative systems (large values of $\varepsilon$). \label{figEREEprod}}
\end{center}
\end{figure}

We calculate and analyse the  entanglement production during the evolution of a 
quantum mechanical dissipative system. Although entanglement entropy plays essential role in the thermalization of isolated quantum systems \cite{PSSV:2011, GE:2016, DKPR:2016}, the central quantity in non-equilibrium systems is not the entropy, but the rate of change of entropy, which is known as entropy production \cite{DM:1984}.  It serves as a measure of the irreversibility of a physical process. 
\\
\indent One can obtain the TFD entanglement entropy production rate (EEPR) by taking the time derivative of the extended non-equilibrium EE (\ref{general EE non-eq case}):
\begin{align}
\nonumber{{\dot S}_A}(t) &=  - {e^{ - \varepsilon {\kern 1pt} t}}\,\left[ {\dot a(t)\,\left( {\log a(t) - 1} \right) - \varepsilon \left( {a(t)\,\log a(t) + \sum\limits_{\mu  = 1}^p {\left( {{S_\mu }(t)\,\tanh \frac{{{K_\mu }}}{4}} \right)} \,\prod\limits_{\alpha  = 1}^p {\coth \frac{{{K_\alpha }}}{4}} } \right)} \right]\\
 &+ (1 - {e^{ - \varepsilon {\kern 1pt} t}})\,\sum\limits_{\mu  = 1}^p {\left( {{{\dot S}_\mu }(t)\,\tanh \frac{{{K_\mu }}}{4}} \right)} \,\prod\limits_{\alpha  = 1}^p {\coth \frac{{{K_\alpha }}}{4}} \,,
\end{align}
where the dot denotes derivative with respect to time $t$. The entanglement entropy production rate $\dot{S}_A(t)$ is shown on figure \ref{figEREEprod} for several values of the dissipation parameter $\varepsilon$. One notices that for short time the EEPR reaches peak value, after which it monotonously decreases with time. The local maximum of $\dot{S}_A(t)$ agrees with the Zeigler's principle of maximum entropy production \cite{NZSH:1963, Zg:1983, ZW:2009}. Furthermore the EEPR is positive quantity which is expected and confirms the statement of the second law of thermodynamics for non-equilibrium systems. One also notes that the peak entropy production is bigger for strongly dissipative systems (large values of $\varepsilon$).
\section{Conclusion}\label{sec conclusion}

In this paper we study the extended entanglement entropy of non-trivial quantized system, namely the closed bosonic string in a homogeneous plane wave geometry. 
\\
\indent The EREE dependence on the inverse scaled temperatures in the ground state of the string has been explicitly shown analytically and graphically in four and five dimensions. The parameter space in these cases is one- and two-dimensional manifold correspondingly. The result shows that EREE increases with increasing temperature, while at very high temperatures (zero inverse scaled temperatures) the entropy diverges. This is also true for the components of the Fisher metric, which is an indication of an instability of the system or a phase transition. To address this problem we have analysed analytically the singularities in the scalar curvature of the metric. We have shown that the Ricci scalar is regular everywhere, including at very high temperatures. This suggests that the considered bosonic string system does not possess any critical points, representing a second-order phase transition, although a first order phase transition is not excluded at high temperature. 
\\
\indent The graphical and analytical study of the scalar curvature showed that there are three geometrically different regions, corresponding to different types of interactions. Near the origin (fig. \ref{curvature t}) the scalar curvature is non-zero and positive, thus defining elliptical geometry on the statistical manifold of thermodynamic parameters. In this case the absolute value of the local maximum of the curvature corresponds to the strongest interaction in the quasi-system. For lower temperatures (fig. \ref{curvature k}) the curvature becomes negative, thus defining a hyperbolic geometry. Here, the strongest interaction is given by the absolute value of the local minimum of the curvature. The value of the scalar curvature on the level curve (fig. \ref{curvature zero}), which separates the different geometric regions, is zero, thus corresponding to a free, non-interacting theory. The limit value of the Ricci scalar at the origin (high temperatures) and at infinity (low temperatures) is also zero, which again leads to a free theory in these cases.
\\
\indent In the one-parameter case the Fisher distance has been derived, which is a measure of dissimilarity between two probability distribution functions. We have shown graphically that in this case the Fisher distance increases monotonously for increasing values of the upper limit of the defining integral. This is interpreted by the theory of fluctuations due to Ruppeiner \cite{GR:1995} in the following manner: the probability of a fluctuation between two equilibrium string states decreases with the increasing of the distance between them.
\\
\indent We have also shown that the general expressions for the entanglement entropy (\ref{standard EE explicit}) and the Fisher information matrix (\ref{Fisher matrix}) work also for other quantum models. Some particular examples were considered, namely BCS-type bosonic systems and XY model in a magnetic field, which is a generalization of the Ising model.
\\
\indent We also manage to derive explicit expression for the entanglement entropy in the non-equilibrium case for a system with dissipation. We showed that the time-dependent entropy increases with time, until it settles at equilibrium (thermalization of the system). The equilibrium value of the entanglement entropy coincides with the thermal equilibrium entanglement entropy derived for a time-independent system. This is also in agreement with the second law of thermodynamics. 
\\
\indent We also depicted one generic behaviour of the entanglement entropy, namely after a transient time which depends on the details of
the initial state of the system, the entanglement entropy goes through a phase of linear or almost linear growth, ${S_A}(t)\sim{\Lambda _A}\,t$, until it settles at the saturation phase. It is also interesting to point out that this kind of behaviour of the entanglement entropy is observed
in the time evolution of various quantum systems that signify quantum chaos \cite{Zurek:1994wd, 1999chao.dyn.11017P, 2000PhRvL..85.3373M, 2002PhRvE..66d5201T}, or in the study of thermalization in some holographic systems \cite{Balasubramanian:2010ce, Balasubramanian:2011ur, Hartman:2013qma, Liu:2013qca}.
\\
\indent The entropy increase rate $\dot{S}_A(t)$ have also been studied for several values of the dissipation parameter $\varepsilon$. We have shown that shortly after the initial moment the entanglement entropy production rate reaches peak value, after which it monotonously decreases with time. The local maximum of $\dot{S}_A(t)$ confirms the Zeigler's principle of maximum entropy production \cite{NZSH:1963, Zg:1983, ZW:2009}. Furthermore the EEPR is positive quantity which is just the statement of the second law of thermodynamics.
\\
\indent Another interesting question is how to reconstruct a parametric family of probability distributions corresponding to the given Fisher metric and to define under what conditions such reconstruction is possible. The Fisher information metric can be straightforwardly calculated once a probability distribution has been chosen. A set of distributions $f(\vec{x},\vec{\theta})$, parametrized by $\vec{\theta}$, forms a statistical manifold. The Riemannian metric on this manifold is the Fisher information metric defined by the following Lebesgue integral:
\begin{equation}\label{distribution rep}
{g_{\mu \nu }}(\vec \theta ) = \int\limits_X {{\cal D}\,f(\vec x,\vec \theta )} \,\frac{{\partial \ln f(\vec x,\vec \theta )}}{{\partial {\theta ^\mu }}}\,\frac{{\partial \ln f(\vec x,\vec \theta )}}{{\partial {\theta ^\nu }}}\,.
\end{equation}
Here $\vec{x}\in X$ is a point from the sample space $X$. It can be proved that the only
Riemannian metric is Fisher metric for which the geometry is invariant
under coordinate transformations of $\vec \theta$ and also under
one-to-one transformations of random variable $\vec x$ \cite{AN:2000, AM:2016}. The Fisher metric is also a solution to the Einstein field equations, which can be useful in finding the corresponding family of probability distributions $f(\vec x,\vec\theta)$. Unfortunately the equations are highly non-linear and cumbersome.  The defining integral (\ref{distribution rep}) only imposes non-trivial constraints on the probability distribution.
\\
\indent {Although this survey is instigated more or less by the fact that
superstring theory on pp-wave backgrounds is exactly solvable, the TFD
framework is powerful enough to treat more complicated supergravity
background solutions. Extending the scope of the present research, next
natural step is to  initiate a more thorough investigation of the
geometric, thermodynamic and information-theoretic aspects of some certain
holographic models.  Such models, for instance, are the $\mathcal{N}=1$ and $\mathcal{N}=2^*$ Pilch-Warner solutions \cite{Pilch:2000ej, Pilch:2000fu, Pilch:2000ue}, Lunin-Maldacena background \cite{Lunin:2005jy}, some recent non-abelian T-dual solutions \cite{Sfetsos:2010uq, Itsios:2012dc, Kelekci:2014ima, Lozano:2011kb, Macpherson:2015tka, Dimov:2015rie} and  their Penrose-G$\ddot{\rm{u}}$ven limit \cite{Penrose1976, Gueven:2000ru} or pp-wave limit \cite{Corrado:2002wi, Brecher:2002ar, Dimov:2015rie}, the latter being easier to study with the techniques used in this paper. Such investigations are expected to shed light on the interplay between
spacetime, global and local properties in holography.

\section*{Acknowledgements}

The authors would like to thank D. Marvakov for the fruitful discussions. This work was partially supported by the Bulgarian NSF grant T02/6 and Sofia University Research Fund grants \textnumero~80-10-116, \textnumero~80-10-118 and \textnumero~80-10-148.




\bibliographystyle{utphys}
\bibliography{eeqdh-refs}

\end{document}